\documentclass[]{article}

\usepackage[english]{babel}
\usepackage{siunitx,bm}
\usepackage{graphicx,rotating,multirow}
\usepackage{amsmath,amssymb,color,mathrsfs,cite,cuted} 
\usepackage{relsize,exscale}

\newcommand{\AJP}{ {\em Am. J. Phys.} }

\newcommand{\AnM}{ {\em Annals Math. }}

\newcommand{\APB}{ {\em Ann. Phys. (Berlin)} }
\newcommand{\APNY}{ {\em Ann. Phys. (N.Y.)} }

\newcommand{\CMP}{ {\em Commun. Math. Phys.} }

\newcommand{\CPL}{ {\em Chem. Phys. Lett.} }
\newcommand{\CRA}{ {\em C. R. Acad. Sci. Ser. A} }
\newcommand{\EJP}{ {\em Eur. J. Phys.} }

\newcommand{\FP}{ {\em Found. Phys.} }

\newcommand{\IJQC}{ {\em Int. J. Quantum Chem.} }
\newcommand{\JAP}{ {\em J. Appl. Phys.} }

\newcommand{\jpa}{ {\em J. Phys. A} }

\newcommand{\JPCM}{ {\em J. Phys. Condens. Matter} }

\newcommand{\JSP}{ {\em J. Stat. Phys.} }

\newcommand{\NJP}{ {\em New J. Phys.} }
\newcommand{\NL}{ {\em Nature (London)} }

\newcommand{\PA}{ {\em Physica A} }

\newcommand{\PLA}{ {\em Phys. Lett. A} }
\newcommand{\PNAS}{ {\em P. Natl. Acad. Sci. USA }}

\newcommand{\PRA}{ {\em Phys. Rev. A} }
\newcommand{\PRB}{ {\em Phys. Rev. B} }

\newcommand{\PRE}{ {\em Phys. Rev. E} }
\newcommand{\PRL}{ {\em Phys. Rev. Lett.} }

\newcommand{\PTP}{ {\em Prog. Theor. Phys. } }

\newcommand{\RMFE}{ {\em Rev. Mex. Fis. E} }
\newcommand{\RMP}{ {\em Rev. Mod. Phys.} }

\begin{document}
\title{Comparative analysis of information measures of the Dirichlet and Neumann two-dimensional quantum dots}
\author{O. Olendski\footnote{{Department of Applied Physics and Astronomy, University of Sharjah, P.O. Box 27272, Sharjah, United Arab Emirates}}}
\maketitle
\begin{abstract}
Analytic representation of both position as well as momentum waveforms of the two-dimensional (2D) circular quantum dots with the Dirichlet and Neumann boundary conditions (BCs) allowed an efficient computation in either space of Shannon $S$, R\'{e}nyi $R(\alpha)$ and Tsallis $T(\alpha)$ entropies, Onicescu energies $O$ and Fisher informations $I$. It is shown that a transition to the 2D geometry lifts the 1D degeneracy of the position components $S_\rho$, $O_\rho$, $R_\rho(\alpha)$. Among many other findings, it is established that the lower limit $\alpha_{TH}$ of the semi-infinite range of the dimensionless R\'{e}nyi/Tsallis coefficient where one-parameter momentum entropies exist is equal to 2/5 for the Dirichlet requirement and 2/3 for the Neumann one. Since their 1D counterparts are $1/4$ and $1/2$, respectively, this simultaneously reveals that this critical value crucially depends  not only on the position BC but the dimensionality of the structure too. As the 2D Neumann threshold $\alpha_{TH}^N$ is greater than one half, its R\'{e}nyi uncertainty relation for the sum of the position and wave vector components $R_\rho(\alpha)+R_\gamma\left(\frac{\alpha}{2\alpha-1}\right)$ is valid in the range $[1/2,2)$ only with its logarithmic divergence at the right edge whereas for all other systems it is defined at any coefficient $\alpha$ not smaller than one half. For both configurations, the lowest-energy level at $\alpha=1/2$ does saturate R\'{e}nyi and Tsallis entropic inequalities. Other properties are discussed and analyzed from the mathematical and physical points of view.

\end{abstract}

\section{Introduction}
Quantum nanostructures are characterized by the large surface-to-volume (for the three-dimensional (3D) dots) or circumference-to-area (2D geometries) ratios. This makes essential a study of the boundary effects, which describe the interaction of the spatially-confined objects with the outer environment. To take into account their influence on the properties of the ultra-small systems, theoretical physics and chemistry supplement the one-particle Shcr\"{o}dinger equation
\begin{equation}\label{Schrodinger1}
-\frac{\hbar^2}{2m^*}{\bm\nabla}_{\bf r}^2\Psi({\bf r})+V({\bf r})\Psi({\bf r})=E\Psi({\bf r})
\end{equation}
for the position wave function $\Psi({\bf r})$ and energy $E$ of the corpuscle of the mass $m^*$ moving in the external potential $V({\bf r})$ by the restriction on $\Psi$ at the interface $\cal S$; for example, a good approximation for semiconductors simply zeroes the waveform at $\cal S$: $\left.\Psi({\bf r})\right|_{\cal S}=0$, what constitutes the Dirichlet boundary condition (BC) whereas for the large class of superconductors \cite{deGennes1} the normal derivative of the order parameter $\Psi$ vanishes at the surface (Neumann BC), $\left.\frac{\partial\Psi}{\partial {\bf n}}\right|_{\cal S}=0$, with $\bf n$ being a unit normal to $\cal S$.

Depending on the BC type, the properties of the structure might differ significantly. \cite{Morse1} This general statement has found its countless confirmations,\cite{Fernandez1,Fernandez2,Fernandez3} one of the most recent analyses \cite{Olendski1} addressing the influence of the Dirichlet and Neumann surface requirements on the R\'{e}nyi \cite{Renyi1,Renyi2} $R$ and Tsallis \cite{Tsallis1} (or, more correctly, Havrda-Charv\'{a}t-Dar\'{o}czy-Tsallis \cite{Havrda1,Daroczy1}) $T$ entropies of the 1D quantum well. These one-parameter functionals are defined in the $d$-dimensional position (subscript $\rho$) and wave vector ($\gamma$) spaces:
\begin{subequations}\label{Renyi1}
\begin{align}\label{Renyi1_R}
R_{\rho_n}(\alpha)&=\frac{1}{1-\alpha}\ln\!\left(\int_{\mathcal{D}_\rho^d}\rho_n^\alpha({\bf r})d{\bf r}\right)\\
\label{Renyi1_K}
R_{\gamma_n}(\alpha)&=\frac{1}{1-\alpha}\ln\!\left(\!\int_{\mathcal{D}_\gamma^d}\gamma_n^\alpha({\bf k})d{\bf k}\right)
\end{align}
\end{subequations}
\begin{subequations}\label{Tsallis1}
\begin{align}\label{Tsallis1_R}
T_{\rho_n}(\alpha)&=\frac{1}{\alpha-1}\left(1-\int_{\mathcal{D}_\rho^d}\rho_n^\alpha({\bf r})d{\bf r}\right)\\
\label{Tsallis1_K}
T_{\gamma_n}(\alpha)&=\frac{1}{\alpha-1}\left(1-\int_{\mathcal{D}_\gamma^d}\gamma_n^\alpha({\bf k})d{\bf k}\right).
\end{align}
\end{subequations}
Here, positive integer index $n$ counts all bound orbitals in the ascending order of their energies and densities $\rho_n({\bf r})=\left|\Psi_n({\bf r})\right|^2$ and $\gamma_n({\bf k})=\left|\Phi_n({\bf k})\right|^2$ are squared magnitudes of the corresponding waveforms, which are orthonormalized:
\begin{equation}\label{OrthoNormality1}
\int_{\mathcal{D}_\rho^d}\Psi_{n'}^\ast({\bf r})\Psi_n({\bf r})d{\bf r}=\int_{\mathcal{D}_\gamma^d}\Phi_{n'}^\ast({\bf k})\Phi_n({\bf k})d{\bf k}=\delta_{nn'},
\end{equation}
with $\delta_{nn'}$ being a Kronecker delta, and related to each other via the Fourier transform:
\begin{subequations}\label{Fourier1}
\begin{align}\label{Fourier1_1}
\Phi_n({\bf k})&=\frac{1}{(2\pi)^{d/2}}\int_{\mathcal{D}_\rho^d}\Psi_n({\bf r})e^{-i{\bf kr}}d{\bf r},\\
\label{Fourier1_2}
\Psi_n({\bf r})&=\frac{1}{(2\pi)^{d/2}}\int_{\mathcal{D}_\gamma^d}\Phi_n({\bf k})e^{i{\bf rk}}d{\bf k}.
\end{align}
\end{subequations}
In all these equations, integrations are carried out over the whole region $\mathcal{D}_\rho^d$ or $\mathcal{D}_\gamma^d$ where the functions $\Psi_n({\bf r})$ or $\Phi_n({\bf k})$ are defined. It was shown \cite{Olendski1}, in particular, that for either BC the dependencies of the R\'{e}nyi position components on the non-negative parameter $\alpha$ are the same for all orbitals but the lowest Neumann one for which the corresponding functional $R$ due to the constant $\Psi_1^N(x)$ is not influenced by the variation of $\alpha$. A crucial difference between the two BCs lies in the fact that the lower limit $\alpha_{TH}$ of the semi-infinite range of the dimensionless R\'{e}nyi/Tsallis coefficient where {\em momentum} entropies exist is equal to one quarter for the Dirichlet requirement and one half for the Neumann one. At $\alpha$ approaching this critical value, the corresponding momentum functionals logarithmically diverge. This distinction causes different behavior of the R\'{e}nyi uncertainty relation \cite{Bialynicki1,Zozor1}
\begin{equation}\label{RenyiUncertainty1}
R_{\rho_n}(\alpha)+R_{\gamma_n}(\beta)\geq-\frac{d}{2}\left(\frac{1}{1-\alpha}\ln\frac{\alpha}{\pi}+\frac{1}{1-\beta}\ln\frac{\beta}{\pi}\right),
\end{equation}
with the positive parameters $\alpha$ and $\beta$ being conjugated as
\begin{equation}\label{RenyiUncertainty2}
\frac{1}{\alpha}+\frac{1}{\beta}=2;
\end{equation}
namely, left-hand side of Equation~\eqref{RenyiUncertainty1} at large $\alpha$ approaches finite Dirichlet values  that are arranged in the increasing order of the index $n$. In the same limit (i.e., at $\beta\rightarrow1/2$), the Neumann sum $\left.\left[R_{\rho_n}^N(\alpha)+R_{\gamma_n}^N(\beta)\right]\right|_{\alpha=\infty}$ diverges due to the similar behavior of the momentum item. Knowledge of the different behavior of the entropic uncertainty relations \cite{Wehner1,Jizba1,Coles1,Toscano1,Hertz1,Wang1} will help in a correct choice of the substances in the design of devices for data compression, quantum cryptography, entanglement witnessing, quantum metrology and other tasks employing correlations between the position and momentum components of the information measures \cite{Coles1,Wang1}. Equation~\eqref{RenyiUncertainty1} is a result of logarithmization of the Sobolev inequality of the Fourier transform \cite{Beckner1}
\begin{equation}\label{Sobolev1}
\left(\frac{\alpha}{\pi}\right)^{d/(4\alpha)}\left[\int_{\mathcal{D}_\rho^d}\rho_n^\alpha({\bf r})d{\bf r}\right]^{1/(2\alpha)}\geq\left(\frac{\beta}{\pi}\right)^{d/(4\beta)}\left[\int_{\mathcal{D}_\gamma^d}\gamma_n^\beta({\bf k})d{\bf k}\right]^{1/(2\beta)},
\end{equation}
for which, in addition to the requirement from Equation~\eqref{RenyiUncertainty2}, an extra constraint
\begin{equation}\label{Sobolev2}
\frac{1}{2}\leq\alpha\leq1
\end{equation}
is imposed and which directly leads to the Tsallis uncertainty relation \cite{Rajagopal1}:
\begin{equation}\label{TsallisInequality1}
\left(\frac{\alpha}{\pi}\right)^{d/(4\alpha)}\!\!\left[1+(1-\alpha)T_{\rho_n}(\alpha)\right]^{1/(2\alpha)}\geq\left(\frac{\beta}{\pi}\right)^{d/(4\beta)}\!\!\left[1+(1-\beta)T_{\gamma_n}(\beta)\right]^{1/(2\beta)}, \end{equation}
which is obviously saturated at $\alpha=\beta=1$ when its either side is equal to $\pi^{-d/4}$. For the 1D Neumann or Dirichlet well, both R\'{e}nyi, Equation~\eqref{RenyiUncertainty1}, as well as Tsallis, Equation~\eqref{TsallisInequality1}, relations are tightened at $\alpha=1/2$ by the lowest orbital \cite{Olendski1} when, e.g., either side of Equation~\eqref{TsallisInequality1} turns to $\Phi_1({\bf 0})$ what appears to be a general property of any magnetic-field-free $d$-dimensional quantum structure \cite{Olendski2,Olendski3}.

Present research expands the previous 1D endeavor \cite{Olendski1} to the comparative analysis of the influence of the Dirichlet and Neumann BCs on the R\'{e}nyi and Tsallis entropies of the 2D circular quantum dot. It is revealed that the lowest thresholds of the range where momentum functionals exist are not only BC-dependent but dimensionality strongly influences them too; namely, the 2D geometry pushes them higher: $\alpha_{TH}^D=2/5$ and $\alpha_{TH}^N=2/3$ where, similar to the 1D well \cite{Olendski1}, Dirichlet value is again smaller than its Neumann counterpart. Since $\alpha_{TH}^N$ is greater than one half, the R\'{e}nyi inequality from Equation~\eqref{RenyiUncertainty1} has its upper boundary at $\alpha=2$ at which its left-hand side diverges whereas for the Dirichlet configuration the corresponding sum is defined for the whole interval of the uncertainty relation $[1/2,+\infty)$. Another specific feature of the 2D symmetry is a non monotonicity of the Neumann entropy $R_\gamma(\alpha)$ in its dependence  on the energy of the level when the growth of the coefficient $\alpha$ leads to the crossings  between R\'{e}nyi momentum functionals of the different orbitals what is also inherited by the corresponding sum from Equation~\eqref{RenyiUncertainty1}. To make the research complete, other measures are considered too; namely, the l'H\^{o}pital's rule degenerates at $\alpha\rightarrow1$ both R\'{e}nyi and Tsallis functionals into the Shannon quantum information entropies \cite{Shannon1}
\begin{subequations}\label{Shannon1}
\begin{align}\label{Shannon1_R}
S_{\rho_n}&=-\int_{\mathcal{D}_\rho^d}\rho_n({\bf r})\ln\rho_n({\bf r})d{\bf r}\\
\label{Shannon1_K}
S_{\gamma_n}&=-\int_{\mathcal{D}_\gamma^d}\gamma_n({\bf k})\ln\gamma_n({\bf k})d{\bf k},
\end{align}
\end{subequations} 
where, as just stated, 
\begin{equation}
S=\lim_{\alpha\rightarrow1}R(\alpha)=\lim_{\alpha\rightarrow1}T(\alpha).
\end{equation}
Its uncertainty relation reads \cite{Bialynicki2,Beckner2}:
\begin{equation}\label{ShannonInequality1}
S_{t_n}\equiv S_{\rho_n}+S_{\gamma_n}\geq d(1+\ln\pi),
\end{equation}
what can be derived as a limiting case of its R\'{e}nyi counterpart, Equation~\eqref{RenyiUncertainty1}. Shannon functionals from Equations~\eqref{Shannon1} measure quantitatively the lack of information about the corresponding characteristics of the object: their smaller (greater) values indicate that we posses more (less) knowledge about corpuscle position, Equation~\eqref{Shannon1_R}, and its wave vector, Equation~\eqref{Shannon1_K}, with the interrelation between them from Equation~\eqref{ShannonInequality1} manifesting that the simultaneous description of both these properties inevitably involves some degree of ignorance, which fundamentally can not be eliminated: the more one learns about the position of the particle, the less knowledge is available about its momentum. From this point of view, the R\'{e}nyi entropy with $\alpha\neq1$ can be construed as a deviation from the equilibrium distribution, which corresponds to the unit coefficient: at the unrestrictedly increasing R\'{e}nyi factor, the events with the highest probabilities are the only ones that determine the corresponding measure whereas the opposite limit of the very small $\alpha$ treats all random events more equally regardless of their actual occurence. As a result, both R\'{e}nyi entropies from Equations~\eqref{Renyi1} monotonically decrease with $\alpha$. Another particular case of the one-parameter functionals are Onicescu energies \cite{Onicescu1}
\begin{subequations}\label{Onicescu1}
\begin{align}\label{Onicescu1_R}
O_{\rho_n}=\int_{\mathcal{D}_\rho^d}\rho_n^2({\bf r})d{\bf r}\\
\label{Onicescu1_K}
O_{\gamma_n}=\int_{\mathcal{D}_\gamma^d}\gamma_n^2({\bf k})d{\bf k},
\end{align}
\end{subequations}
which yield the numbers that categorize deviations of the probability distributions from the uniform ones. From comparison between Equations~\eqref{Renyi1}, \eqref{Tsallis1} and \eqref{Onicescu1}, it immediately follows that $O_{\rho,\gamma}=e^{-R_{\rho,\gamma}(2)}=1-T_{\rho,\gamma}(2)$. Just second-order many-body R\'{e}nyi entanglement entropy $R(2)$ was measured in recent experiments on the Bose-Einstein condensates of the interacting atoms and ions \cite{Islam1,Kaufman1,Brydges1}. Even though both Shannon and Onicescu quantities are special cases of the more general R\'{e}nyi/Tsallis dependence, they are of interest in their own right what justifies their separate consideration. In addition, Fisher informations \cite{Fisher1,Frieden1}
\begin{subequations}\label{Fisher1}
\begin{align}\label{Fisher1_R}
I_{\rho_n}&=\int_{\mathcal{D}_\rho^d}\rho_n({\bf r})\left|{\bm\nabla}_{\bf r}\ln\rho_n({\bf r})\right|^2d{\bf r}=\int_{\mathcal{D}_\rho^d}\frac{\left|{\bm\nabla}_{\bf r}\rho_n({\bf r})\right|^2}{\rho_n({\bf r})}d{\bf r}\\
\label{Fisher1_K}
I_{\gamma_n}&=\int_{\mathcal{D}_\gamma^d}\gamma_n({\bf k})\left|{\bm\nabla}_{\bf k}\ln\gamma_n({\bf k})\right|^2d{\bf k}=\int_{\mathcal{D}_\gamma^d}\frac{\left|{\bm\nabla}_{\bf k}\gamma_n({\bf k})\right|^2}{\gamma_n({\bf k})}\,d{\bf k}
\end{align}
\end{subequations}
are addressed too, which - due to the presence of the gradients - provide a quantitative estimation of the oscillating structure of each probability distribution. In one of its applications, the position Fisher information defines the kinetic energy of the many-particle system what allows to reformulate the quantum mechanical variation principle as a principle of minimal information \cite{Sears1}. Expressions~\eqref{Onicescu1} and \eqref{Fisher1} show that the last two measures can not be negative whereas Shannon and R\'{e}nyi entropies for the continuous probablitity distributions might fall below zero with the Tsallis functionals containing the items of the different dimensionalities, what will be discussed below. Let us also mention that for the Onicescu energies and Fisher informations there are no universal uncertainty relations linking their position and momentum components.

Exposition starts from addressing, in Section~\ref{sec_WF1}, the wave functions $\Psi({\bf r})$ and $\Phi({\bf k})$, which are the building blocks of all measures. Even though position dependencies for both BCs are fairly well known from the electromagnetic theory of the metallic guiding structures \cite{Harrington1}, our consideration focuses on some not discussed before properties pertinent to quantum information theory. Moreover, previous analysis of the momentum waveforms was limited to the azimuthally symmetric levels of the Dirichlet disc only \cite{Song1} whereas below with the help of the Bessel functions \cite{Abramowitz1} exact quite simple expressions are obtained for all orbitals of both geometries what greatly facilitates the understanding of the wave vector components of the functionals $S$, $O$, $I$, $R$ and $T$. As a mathematical byproduct, expressions for some improper integrals involving Bessel functions are derived. Section~\ref{sec_Shannon} is devoted to the study of the parameter-free measures where, in addition to the numerical computation, a special care is paid to the analytic properties in the limiting cases and comparison of the two edge requirements. A comparative study of the two BCs demonstrates, in particular, a decreasing difference between the corresponding measures at the larger principal $n$ and azimuthal $m$ indexes what is explained by the lesser sensitivity of the higher-energy quantum states to the interface requirement. Mathematical results for $S$, $I$ and $O$ are explained from physical point of view; for instance, unrestricted decrease (increase) at $|m|$ or $n$ tending to infinity of the position component of the Shannon entropy (Onicescu energy) is due to the transition from the quantum regime to the quasi-classical description. The same line of research is retained in Section~\ref{sec_Renyi}, which sheds numerical and analytic light on the R\'{e}nyi and Tsallis entropies. Section~\ref{sec_Conclusions} summarizes the findings and discusses their possible extensions, including expressions for the $d$-dimensional generalization of $\alpha_{TH}^D$ and $\alpha_{TH}^N$.

\section{Wave functions}\label{sec_WF1}
For the analysis of the quantum motion inside the circle of radius $a$, one needs to use the polar coordinates ${\bf r}\equiv(r,\varphi_{\bf r})$ with its origin in the middle of the structure. Then, the domain of the Hamiltonian
\begin{equation}\label{Hamiltonian1}
\widehat{H}=-\frac{\hbar^2}{2m^*}{\bm\nabla}_{\bf r}^2=-\frac{\hbar^2}{2m^*}\left(\frac{\partial^2}{\partial r^2}+\frac{1}{r}\frac{\partial}{\partial r}+\frac{1}{r^2}\frac{\partial^2}{\partial\varphi_{\bf r}^2}\right)
\end{equation}
consists of the differentiable functions $\Psi^{D,N}(r,\varphi_{\bf r})$ defined inside the dot with the corresponding BC at the circumference $r=a$:
\begin{equation}\label{Domain1}
\mathcal{D}_\rho^2\left(-\frac{\hbar^2}{2m^*}{\bm\nabla}_{\bf r}^2\right)=\left\{\Psi,{\bm\nabla}_{\bf r}^2\Psi\in\mathcal{L}^2(0\leq r<a,0\leq\varphi_{\bf r}<2\pi),\left\{\begin{array}{c}
\Psi^D(a,\varphi_{\bf r})=0\\
\left.\frac{\partial\Psi^N(r,\varphi_{\bf r})}{\partial r}\right|_{r=a}=0
\end{array}\right\}\right\},
\end{equation}
Due to the rotational symmetry, variables are separated:
\begin{equation}\label{Separation1}
\Psi_{nm}(r,\varphi_{\bf r})=\frac{1}{\sqrt{2\pi}}e^{im\varphi_{\bf r}}{\cal R}_{nm}(r),
\end{equation}
where $n=1,2\ldots$ and $m=0,\pm1,\ldots$ are principal and azimuthal indices, respectively. To guarantee the orthonormalization, Equation~\eqref{OrthoNormality1}, the radial dependence ${\cal R}_{nm}(r)$ should satisfy:
\begin{equation}\label{OrthoNormality2}
\int_0^ar{\cal R}_{nm}(r){\cal R}_{n'm}(r)dr=\delta_{nn'}.
\end{equation}

Then, the Fourier transform, Equation~\eqref{Fourier1_1}, yields for the momentum function $\Phi_{nm}(k,\varphi_{\bf k})$ that is represented in the wave vector polar coordinates ${\bf k}\equiv(k,\varphi_{\bf k})$, $0\leq k<\infty$, $0\leq\varphi_{\bf k}<2\pi$:
\begin{equation}\label{PositionMomentumRelation2}
\Phi_{nm}(k,\varphi_{\bf k})=\frac{1}{(2\pi)^{3/2}}\int_0^a\!\!drr{\cal R}_{nm}(r)\int_0^{2\pi}\!\!d\varphi_{\bf r}e^{i[m\varphi_{\bf r}-kr\cos(\varphi_{\bf r}-\varphi_{\bf k})]}.
\end{equation}
Angular integration in Equation~\eqref{PositionMomentumRelation2} is carried out with the help of the identity \cite{Morse2}:
\begin{equation}\label{PolarIdentity1}
\int_0^{2\pi}e^{i[m\theta_1-z\cos(\theta_1-\theta_2)]}d\theta_1=(-i)^m2\pi J_m(z)e^{im\theta_2},
\end{equation}
where $J_\nu(z)$ is $\nu$-th order Bessel function of the first kind \cite{Abramowitz1}. This equation means that the momentum dependence is separated into the angular and radial parts too:
\begin{subequations}\label{Separation2}
\begin{align}\label{Separation2_1}
\Phi_{nm}(k,\varphi_{\bf k})&=\frac{(-i)^{|m|}}{\sqrt{2\pi}}e^{im\varphi_{\bf k}}{\cal K}_{nm}(k),
\intertext{with}\label{Separation2_2}
{\cal K}_{nm}(k)&=\int_0^ar{\cal R}_{nm}(r)J_{|m|}(kr)dr,
\intertext{where, due to the condition from Equation~\eqref{OrthoNormality1},}
\label{Separation2_3}
&\int_0^\infty k{\cal K}_{nm}(k){\cal K}_{n'm}(k)dk=\delta_{nn'}.
\end{align}
\end{subequations}

Depending on the BC type, position ${\cal R}_{nm}(r)$ and, accordingly, momentum ${\cal K}_{nm}(k)$  functions  take different forms.

\subsection{Dirichlet BC}\label{sec_WaveFunctionsDirichlet}
Upon substituting the total position wave function, Equation~\eqref{Separation1}, into the Schr\"{o}dinger equation~\eqref{Schrodinger1} with the Hamiltonian from Equation~\eqref{Hamiltonian1}, its radial part ${\cal R}_{nm}(r)$ that satisfies the Bessel equation \cite{Abramowitz1} takes the form ${\cal R}_{nm}(r)=N_{n|m|}J_{|m|}\left(\sqrt{\frac{2m^*E}{\hbar^2}}\,r\right)$, where $N_{n|m|}$ is a normalization constant and where the second solution of the Bessel equation $Y_{|m|}\left(\sqrt{\frac{2m^*E}{\hbar^2}}\,r\right)$ has been dropped due to its divergence at the origin \cite{Abramowitz1}. For the dependence that  vanishes at the rim,
${\cal R}_{nm}^D(a)=0$, the discrete energy spectrum turns to:
\begin{equation}\label{DirichletSpectrum1}
E_{nm}^D=\frac{\hbar^2}{2m^*a^2}j_{|m|n}^2,
\end{equation}
with $j_{\nu s}$ being $s$th root of the $\nu$th order Bessel function \cite{Abramowitz1}, $J_\nu(j_{\nu s})=0$, whereas the orthonormalized, Equation~\eqref{OrthoNormality2}, radial functions become
\begin{equation}\label{RadialFunctionDIrichlet1}
{\cal R}_{nm}^D(r)=\frac{2^{1/2}}{aJ_{|m|+1}(j_{|m|n})}J_{|m|}\left(j_{|m|n}\frac{r}{a}\right),
\end{equation}
what yields the following expression of the total position waveforms:
\begin{subequations}\label{DirichletFunction1}
\begin{align}\label{DirichletFunction1_Position}
\Psi_{nm}^D\left(r,\varphi_{\bf r}\right)&=\frac{1}{\pi^{1/2}aJ_{|m|+1}(j_{|m|n})}J_{|m|}\left(j_{|m|n}\frac{r}{a}\right)\,e^{im\varphi_{\bf r}}.
\intertext{By subsequent analytic evaluation of the radial quadrature from Equation~\eqref{Separation2_2} with the help of known tables \cite{Gradshteyn1,Prudnikov2}, one arrives ultimately at}
\label{DirichletFunction1_Momentum}
\Phi_{nm}^D\left(k,\varphi_{\bf k}\right)&=a\frac{(-i)^{|m|}}{\pi^{1/2}}\frac{j_{|m|n}}{j_{|m|n}^2-(ak)^2}J_{|m|}(ak)e^{im\varphi_{\bf k}}.
\end{align}
\end{subequations}
Orthonormalization \eqref{Separation2_3} results in the following integral:
\begin{equation}\label{OrthonormalizationDirichlet1}
\int_0^\infty\frac{\xi}{\left(j_{|m|n}^2-\xi^2\right)\left(j_{|m|n'}^2-\xi^2\right)}J_{|m|}^2(\xi)d\xi=\frac{1}{2j_{|m|n}^2}\delta_{nn'},
\end{equation}
which is absent in known literature \cite{Abramowitz1,Gradshteyn1,Prudnikov2,Prudnikov3,Brychkov1}.

Knowledge of the analytic representation of the position and momentum wave functions and, accordingly, of their associated
densities
\begin{subequations}\label{DirichletDensities1}
\begin{eqnarray}\label{DirichletDensities1_R}
\rho_{nm}^D({\bf r})&\equiv\rho_{nm}^D(r)=&\frac{1}{\pi a^2}\left[\frac{J_{|m|}(j_{|m|n}r/a)}{J_{|m|+1}(j_{|m|n})}\right]^2\\
\label{DirichletDensities1_P}
\gamma_{nm}^D({\bf k})&\equiv\gamma_{nm}^D(k)=&\frac{a^2}{\pi}\left[\frac{j_{|m|n}}{j_{|m|n}^2-(ak)^2}J_{|m|}(ak)\right]^2
\end{eqnarray}
\end{subequations}
paves the way for an efficient computation of all other characteristics in expressions that they enter, Equations~\eqref{Renyi1}, \eqref{Tsallis1}, \eqref{Shannon1}, \eqref{Onicescu1} and \eqref{Fisher1}, what will be the subject of the subsequent sections. Right now, let us only point out that at the large wave vectors the Dirichlet momentum density decays to zero as
\begin{equation}\label{AsymptoteDirichlet1}
\gamma_{nm}^D(k)\sim\frac{1}{k^5},\quad k\rightarrow\infty.
\end{equation}
\begin{figure}
\centering
\includegraphics[width=\columnwidth]{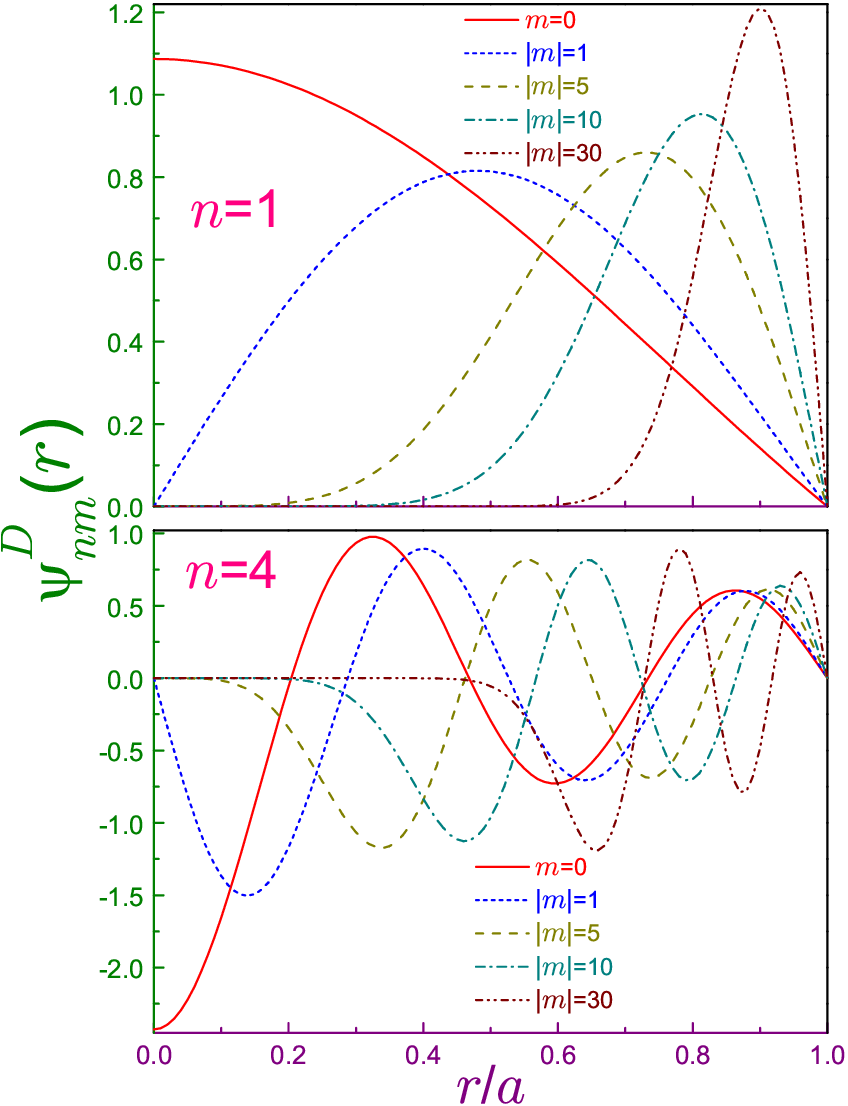}
\caption{\label{Fig_FunctionsDirichletR1}
Dirichlet position functions $\psi_{nm}^D(r)$ from Equation~\eqref{DirichletWaveForm2} in terms of dimensionless distance $r/a$ for $n=1$ (upper panel) and $n=4$ (lower window) where solid lines are for $|m|=0$, dotted curves -- for $|m|=1$, dashed ones are for $|m|=5$, dash-dotted lines -- for $|m|=10$, and dash-dot-dotted curves -- for $|m|=30$. Note different vertical ranges and scales of the subplots what is the case for Figures~\ref{Fig_FunctionsDirichletP1} -- \ref{Fig_FunctionsNeumannP1} too.}
\end{figure}
\begin{table}[bt]
\caption{Dirichlet root-mean-square radius $\overline{r}_{nm}^D$ (in units of the dot radius $a$), Equation~\eqref{MeanRadiusDirichlet1}, for several radial $n$ and azimuthal $m$ quantum numbers}
\centering
\begin{tabular}{|c|c|c|c|c|c|c|c|c|c|}
\hline
\multirow{2}{1em}{$|m|$}& \multicolumn{9}{c|}{Principal quantum number $n$}\\ 
\cline{2-10}
&$n=1$&$n=2$&$n=3$&$n=4$&$n=5$&$n=10$&$n=20$&$n=50$&$n=100$\\
\cline{1-10}
0&0.4670&0.5581&0.5696&0.5732&0.5748&0.5767&0.5772&0.5773&0.5773\\
1&0.5774&0.5774&0.5774&0.5774&0.5774&0.5774&0.5774&0.5774&0.5774\\
2&0.6397&0.6013&0.5900&0.5852&0.5827&0.5789&0.5778&0.5774&0.5774\\
3&0.6814&0.6239&0.6040&0.5946&0.5895&0.5811&0.5784&0.5775&0.5774\\
4&0.7120&0.6442&0.6179&0.6046&0.5970&0.5837&0.5792&0.5777&0.5774\\
5&0.7357&0.6621&0.6311&0.6146&0.6048&0.5868&0.5802&0.5779&0.5775\\
10&0.8052&0.7263&0.6849&0.6594&0.6422&0.6046&0.5869&0.5793&0.5779\\
20&0.8632&0.7935&0.7507&0.7209&0.6989&0.6409&0.6044&0.5839&0.5793\\
50&0.9187&0.8693&0.8348&0.8081&0.7864&0.7175&0.6562&0.6042&0.5868\\
100&0.9466&0.9116&0.8856&0.8646&0.8467&0.7841&0.7164&0.6397&0.6041\\
200&0.9655&0.9417&0.9234&0.9081&0.8947&0.8447&0.7829&0.6950&0.6396\\
500&0.9809&0.9671&0.9563&0.9470&0.9387&0.9060&0.8608&0.7822&0.7154\\
1000&0.9878&0.9790&0.9719&0.9657&0.9602&0.9378&0.9053&0.8431&0.7820\\
\hline
\end{tabular}
\label{Table_MeanRadiusDirichlet}
\end{table}

Figure~\ref{Fig_FunctionsDirichletR1} depicts dimensionless real functions
\begin{equation}\label{DirichletWaveForm2}
\psi_{nm}(r)=a\Psi_{nm}(r,\varphi_{\bf r})e^{-im\varphi_{\bf r}},
\end{equation}
which do not depend on the sign of $m$, for several values of the principal and magnetic quantum indexes. Rotationally symmetric orbitals are the only ones at which the value of the probability in the middle of the dot is not zero:
\begin{equation}\label{Equation1R_Dirichlet}
\rho_{nm}^D(0)=\frac{1}{\pi a^2}\frac{1}{J_0'^{\,2}\!\left(j_{0n}\right)}\delta_{m0}.
\end{equation}
Since the number of the extrema is equal to the radial index, the frequency of oscillations does increase for the higher $n$, which - due to the presence of the gradient in Equation~\eqref{Fisher1_R} - results in higher values of the position Fisher information, see Table~\ref{Table_FisherDirichlet}. Next, as expected, the states with the greater $|m|$ are shifted further from the origin and thus, accumulate stronger at the edge with the absolute value of the nearest extremum increasing as the large magnetic number grows. To quantify this phenomenon, one notices that the extrema of ${\cal R}_{nm}^D(r)$ and, accordingly, the maxima of $\rho_{nm}^D(r)$ are located at
\begin{subequations}\label{ExtremaDirichlet1}
\begin{align}\label{ExtremaDirichlet1_Pos}
r_{nmn'_{max}}^D=&\frac{j_{|m|n'}'}{j_{|m|n}}a,\quad n'=1,2,\ldots,n,
\intertext{where $j_{\nu s}'$ is $s$-th zero of the derivative of the $\nu$th order Bessel function \cite{Abramowitz1}, $J_\nu'(j_{\nu s}')=0$. The magnitude at the maximum is:}
\label{ExtremaDirichlet1_Magn}
\rho_{nmn'_{max}}^D=&\frac{1}{\pi a^2}\left[\frac{J_{|m|}(j_{|m|n'}')}{J_{|m|}'(j_{|m|n})}\right]^2
\end{align}
\end{subequations}
with the first peak, $n'=1$, being the largest one. Taking as a representative example the lowest band, $n=1$, from the properties of the Bessel functions \cite{Abramowitz1} one finds that at the huge magnetic index the only maximum shifts closer to the surface and its magnitude diverges:
\begin{subequations}\label{ExtremaDirichlet2}
\begin{align}\label{ExtremaDirichlet2_Pos}
r_{nmn'_{max}}^D\rightarrow&a,\quad|m|\rightarrow\infty\\
\label{ExtremaDirichlet2_Magn}
\rho_{1m1_{max}}^D\rightarrow&\frac{0.3394}{\pi a^2}|m|^{2/3},\quad|m|\rightarrow\infty.
\end{align}
\end{subequations}
This is clearly seen in the upper window. For the higher lying bands, these asymptotes (with different numerical coefficients in Equation~\eqref{ExtremaDirichlet2_Magn} for different $n$) are reached at the greater $|m|$. The same conclusion can be arrived at if one considers a root-mean-square radius $\overline{r}=\sqrt{\int_{{\cal D}_\rho^d}r^2\rho({\bf r})d{\bf r}}$, what for the 2D Dirichlet disc becomes
\begin{equation}\label{MeanRadiusDirichlet1}
\overline{r}_{nm}^D=\frac{a}{\left|J_{|m|+1}(j_{|m|n})\right|}\sqrt{2\int_0^1z^3J_{|m|}^2(j_{mn}z)dz}.
\end{equation}
For $|m|=1$, an analytic calculation \cite{Prudnikov2} reveals that it is independent of the principal index, $\overline{r}_{n,\pm1}^D=a/3^{1/2}=a\,0.57735\ldots$. Table~\ref{Table_MeanRadiusDirichlet} shows that for the higher bands its convergence to the boundary with the growing $|m|$ takes place at the greater absolute value of the magnetic index. It also follows that $\overline{r}_{n0}^D$ $\left(\overline{r}_{n,m>1}^D\right)$ is an increasing (decreasing) sequence of the principal quantum number which in the case of the huge $n$ approaches from below (above) the $|m|=1$ limit with this asymptote for the higher $|m|$ being reached at the greater $n$. Let us also mention that the mean radius $\overline{\mathtt{r}}=\int_{{\cal D}_\rho^d}r\rho({\bf r})d{\bf r}$ at the huge magnetic index qualitatively tends to the rim very similar to its root-mean-square counterpart from Eq.~\eqref{MeanRadiusDirichlet1} whereas the unrestrictedly  increasing principal number brings $\overline{\mathtt{r}}_{nm}^D$ closer and closer to $a/2$. To save space, its detailed behavior is not shown here.

\begin{figure}
\centering
\includegraphics[width=\columnwidth]{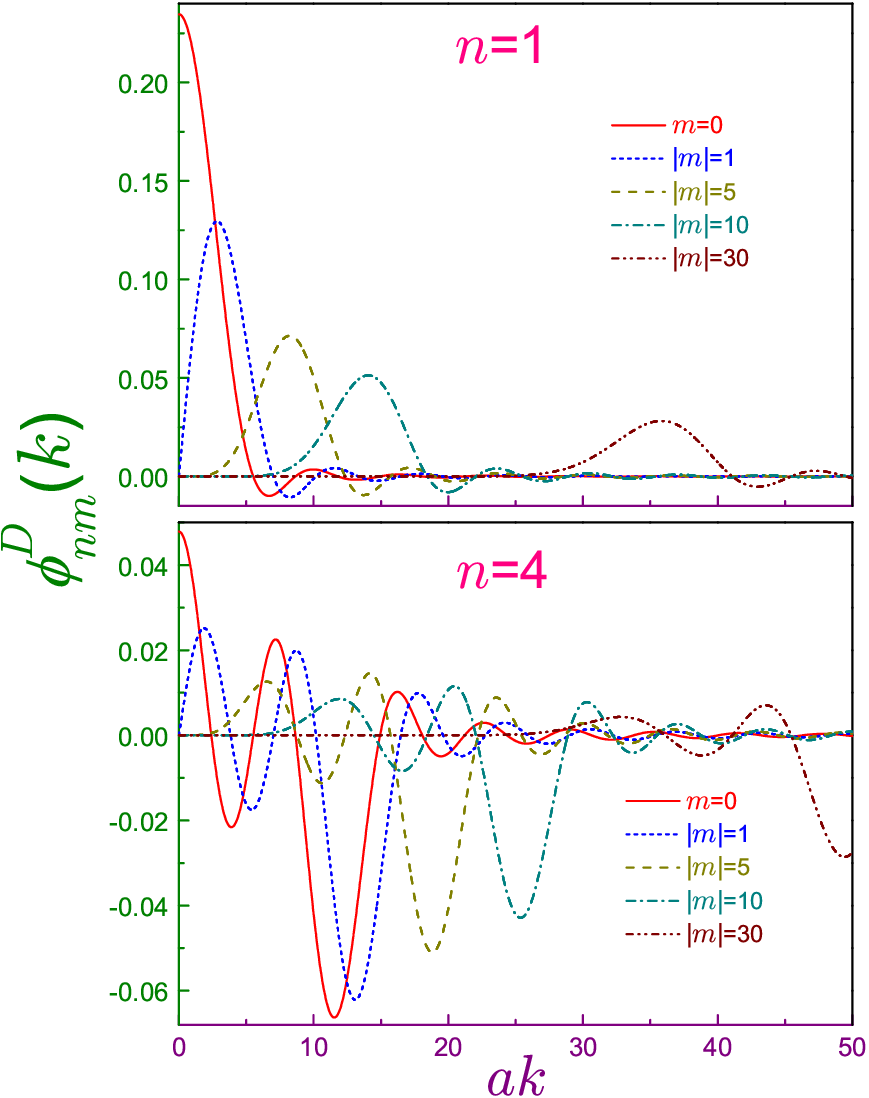}
\caption{\label{Fig_FunctionsDirichletP1}
Dirichlet momentum functions $\phi_{nm}^D(k)$ from Equation~\eqref{MomentumDirichletFunctions2} in terms of dimensionless wave vector $ak$. The same convention as in Figure~\ref{Fig_FunctionsDirichletR1} is adopted.}
\end{figure}

Figure~\ref{Fig_FunctionsDirichletP1} depicts several real dimensionless momentum waveforms
\begin{equation}\label{MomentumDirichletFunctions2}
\phi_{nm}(k)=\frac{i^{|m|}}{a}\Phi_{nm}(k,\varphi_{\bf k})e^{-im\varphi_{\bf k}}
\end{equation}
for the Dirichlet BC. From the properties of the Bessel functions it follows that, similar to the position component, only the $m=0$ levels do not suppress the function $\Phi_{nm}(k,\varphi_{\bf k})$ at the origin $k=0$:
\begin{equation}\label{Equation1G_Dirichlet}
\gamma_{nm}^D(0)=\frac{1}{\pi}\left(\frac{a}{j_{|m|n}}\right)^2\delta_{m0},
\end{equation}
what for the ground orbital, $n=1$, is its global maximum. Nonvanishing probability of the zero wave vector is a peculiar feature of the rotationally symmetric Dirichlet states, which is inherited from the corresponding property of the position dependencies, Eq.~\eqref{Equation1R_Dirichlet}. Also, it is seen that, in particular, the states with the larger magnetic index are shifted to the greater momenta $k$. Dirichlet momentum density $\gamma_{nm}^D(k)$ possesses an infinite number of tops and zero-valued valleys and reaches its largest extremum at the $n$-th peak, which in the case of huge $|m|$ and/or $n$ is located at
\begin{subequations}\label{DirichletMomentumMax1}
\begin{align}\label{DirichletMomentumMax1_Location}
k_{nm_{max}}^D&=\frac{j_{|m|n}}{a}\left(1-\frac{3}{j_{|m|n}^2-m^2-5}\right),\quad |m|\,\,{\rm and/or}\,\,n\rightarrow\infty,
\intertext{with its magnitude being}\label{DirichletMomentumMax1_Magnitude}
\gamma_{nm_{max}}^D&=\frac{a^2}{4\pi}J_{|m|}'^{\,2}\!\left(j_{|m|n}\right)\!\left(1\!+\!\frac{6}{j_{|m|n}^2-m^2-5}\!\right),\,|m|\,\,{\rm and/or}\,\,n\rightarrow\infty.
\end{align}
\end{subequations}
Expression $J_{|m|}'^{\,2}\!\left(j_{|m|n}\right)$ in the limit $|m|\rightarrow\infty$ approaches zero as $|m|^{-4/3}$ \cite{Abramowitz1}. If, similar to the position space, one defines a root-mean-square wave vector $\overline{k}$ as a square root of its second-order moment, $\overline{k}=\sqrt{\left\langle k^2\right\rangle}$, where
\begin{align}
\label{Kmoment1}
\left\langle k^j\right\rangle=\int_{{\cal D}_\gamma^d}k^j\gamma({\bf k})d{\bf k},\quad j=0,1,\ldots,
\intertext{then for the Dirichlet dot it at any $m$ and $n$ is equal (in units of $1/a$) to the corresponding Bessel zero:}
\label{MeanWaveVectorDirichlet1}
\overline{k}_{nm}^D=\frac{j_{|m|n}}{a},
\intertext{what means that}
\label{MeanWaveVectorDirichlet2}
\int_0^\infty \frac{\xi^3}{\left(j_{|m|n}^2-\xi^2\right)^2}J_{|m|}^2(\xi)d\xi=\frac{1}{2}.
\end{align}

It is important to state that, here, the magnitude of the total wave vector $k$, even though being a $c$-number, in the momentum representation is considered an operator and, accordingly, expression~\eqref{Kmoment1} with $j=1$ and $j=2$ is its and its square expectation values just in this rendering. Then, Equation~\eqref{MeanWaveVectorDirichlet1} follows straightforwardly from the Dirichlet spectrum, Equation~\eqref{DirichletSpectrum1}, and energy dependence on the wave vector, $E=\frac{\hbar^2k^2}{2m^*}$.
On the other hand, in the position representation, the radial momentum is a differential operator:
\begin{subequations}\label{MomentumOperator1}
\begin{align}\label{MomentumOperator1_1}
\widehat{k}_r&=-i\left(\frac{\partial}{\partial r}+\frac{d-1}{2r}\right).
\intertext{First, it is elementary to see that for the spatially confined domains none of the solutions, Equation~\eqref{DirichletFunction1_Position}, of the Schr\"{o}dinger equation with \textit{real} $r$-dependence is its eigenstate what is not surprising as, semiclassically, the particle after hitting the wall changes its direction of motion and, accordingly, radial momentum is not conserved with its averaging over the whole trajectory being zero, $\left\langle \widehat{k}_r\right\rangle=0$. Second, its square}\label{MomentumOperator1_2}
\widehat{k}_r^2&=-\left[\frac{\partial^2}{\partial r^2}+\frac{d-1}{r}\frac{\partial}{\partial r}+\frac{(d-1)(d-3)}{4r^2}\right]
\end{align}
\end{subequations}
times $\hbar^2/(2m^*)$, when added to the square of the angular momentum operator $\widehat{{\bf L}}^2$ (divided by $2m^*r^2$), does not coincide with the Hamiltonian  only for the 2D systems \cite{Paz1,Fujikawa1}:
\begin{equation}\label{Hamiltonian2}
\widehat{H}=\frac{\hbar^2\widehat{k}_r^2}{2m^*}+\frac{\widehat{{\bf L}}^2}{2m^*r^2}+\frac{\hbar^2}{2m^*}\frac{(d-1)(d-3)}{4r^2}.
\end{equation}
From this, one gets at $d=2$:
\begin{equation}\label{MomentumOperator2}
\left\langle \widehat{k}_r^2\right\rangle=\frac{2m^*}{\hbar^2}E-\left(m^2-\frac{1}{4}\right)\left\langle\frac{1}{r^2}\right\rangle.
\end{equation}
For the Dirichlet BC with  the position functions from Equation~\eqref{DirichletFunction1_Position}, one finds the expectation value of the inverse square radius to be
\begin{equation}\label{DirichletInverseRadius1}
\left\langle\frac{1}{r^2}\right\rangle_{\!\!nm}^{\!\!D}=\frac{1}{a^2}\frac{1}{|m|J_{|m|}'^{\,2}(j_{|m|n})}\left[1-J_0^2(j_{|m|n})-2\sum_{k=1}^{|m|-1}J_k^2(j_{|m|n})\right],
\end{equation}
and then
\begin{equation}\label{MomentumOperator3}
\left\langle \widehat{k}_r^2\right\rangle_{\!\!nm}^{\!\!D}=\frac{1}{a^2}\left(j_{|m|n}^2-\frac{1}{J_{|m|}'^{\,2}(j_{|m|n})}\left(|m|-\frac{1}{4|m|}\right)\left[1-J_0^2(j_{|m|n})-2\sum_{k=1}^{|m|-1}J_k^2(j_{|m|n})\right]\right),
\end{equation}
which, due to the singularity at the origin of the last item in the right-hand side of Equation~\eqref{MomentumOperator1_2}, diverges for the azimuthally symmetric orbitals, $m=0$. Accordingly, as the infinity is greater than any finite number, this makes meaningless the 2D Heisenberg uncertainty relation \cite{Bracher1}  for these Dirichlet states
\begin{equation}\label{Heisenberg1}
\overline{r}\,\overline{k}\geq|m|+1,
\end{equation}
if the momentum expectation value is calculated in the position representation but Equation~\eqref{Heisenberg1} still provides some bound if $\overline{k}$ is evaluated in the wave vector rendering, Equation~\eqref{Kmoment1}. This example sheds another doubt on the definition of the radial linear momentum operator $\widehat{p}_r=\hbar\widehat{k}_r$ in the form from Equation~\eqref{MomentumOperator1_1}. Despite the fact that it does obey a correct commutation relation with the absolute value of the radius-vector,  $\left[r,\widehat{p}_r\right]=i\hbar$, its very definition and meaning still remain a matter of controversy; e.g., for $d=3$, P. A. M. Dirac claims that $\widehat{p}_r$ "is real and is a true momentum conjugate to $r$"  \cite{Dirac1}. However, R. H. Dicke and J. P. Wittke retort that $\widehat{p}_r$ "is not the $r$-component of the particle momentum" \cite{Dicke1} with A. Messiah adding that "$\widehat{p}_r$ is Hermitian but is not an observable" \cite{Messiah1} as its eigenvalue problem, according to him, has no solution. Several authors tried to prove that $\widehat{p}_r$ is not Hermitian \cite{Liboff1,Levin1,Twamley1,Paz2}. In addition to the just mentioned sources \cite{Paz1,Fujikawa1,Dirac1,Dicke1,Messiah1,Liboff1,Levin1,Twamley1,Paz2} and literature cited therein, we recommend to the reader References~\cite{Essen1,Domingos1,Roy1} where more info on the subject is provided including historic references dating back to the early days of the wave mechanics. In a most recent development, the experiment employed light modulators for studying the 2D spatially entangled photon pairs generated by spontaneous parametric down-conversion \cite{Chen1}. Researchers used the fact that in the infinite 2D domain the radial momentum eigenstate with the continuous eigenvalue $p$ reads in the position representation as $\Psi_p(r)=\frac{1}{2\pi\hbar}\frac{\exp(ipr/\hbar)}{r^{1/2}}$. Measurements revealed correlations between the radial position and radial momentum of the two separated photons, which violate the Heisenberg uncertainty relation what convincingly proves the entangled nature of the radial structure of two-photon wave functions. Provided above is an example of shortcomings of the operator $\widehat{k}_r$ on the the spatially confined 2D region, which supplements the previous analysis for the unbound position space \cite{Paz1,Fujikawa1,Dirac1,Dicke1,Messiah1,Liboff1,Levin1,Twamley1,Paz2,Essen1,Domingos1,Roy1}.

Wrapping up a discussion on the Dirichlet momentum waveform, let us point out that its fading and flattening on increasing $|m|$  can be also shown explicitly from its analytic representation, Equation~\eqref{DirichletFunction1_Momentum}, and known asymptotic formula for the Bessel function with large index, $J_\nu(z)\rightarrow(2\pi\nu)^{-1/2}[ez/(2\nu)]^\nu,\quad\nu\rightarrow\infty$ \cite{Abramowitz1}.

Thus, summing up this part of research, it has  been shown that, in addition to the position part, momentum wave functions for the arbitrary principal $n$ and magnetic $m$ indexes are also represented {\em analytically}. As shown in the next subsection, one arrives at the same conclusion for the Neumann BC too. Such a consideration allows us to find some properties of both densities what will be used in the analysis of the quantum-information measures.

\subsection{Neumann BC}\label{sec_WaveFunctionsNeumann}
When the derivative of the position radial dependence turns to zero at the boundary, $\left.\frac{d}{dr}{\cal R}_{nm}^N(r)\right|_{r=a}=0$, the energy spectrum reads:
\begin{equation}\label{NeumannSpectrum1}
E_{nm}^N=\frac{\hbar^2}{2m^*a^2}j_{|m|n}'^{\,2},
\end{equation}
and the associated waveforms
\begin{subequations}\label{NeumannFunction1}
\begin{align}\label{NeumannPositionFunction1}
\Psi_{nm}^N(r,\varphi_{\bf r})&=\frac{1}{\pi^{1/2}a}
\frac{j_{|m|n}'}{\left(j_{|m|n}'^{\,2}-m^2\right)^{1/2}}\frac{J_{|m|}\left(j_{|m|n}'\frac{r}{a}\right)}{J_{|m|}\left(j_{|m|n}'\right)}\,e^{im\varphi_{\bf r}}\\
\label{NeumannMomentumFunction1}
\Phi_{nm}^N(k,\varphi_{\bf k})&=(-i)^{|m|}\frac{a}{\pi^{1/2}}\frac{j_{|m|n}'}{\left(j_{|m|n}'^{\,2}-m^2\right)^{1/2}}\frac{ak}{j_{|m|n}'^{\,2}-(ak)^2}J_{|m|}'(ak)\,e^{im\varphi_{\bf k}}
\end{align}
\end{subequations}
define the corresponding densities as:
\begin{subequations}\label{NeumannDensities1}
\begin{align}\label{NeumannDensities1_R}
\rho_{nm}^N({\bf r})&\equiv\rho_{nm}^N(r)=\frac{1}{\pi a^2}\frac{j_{|m|n}'^{\,2}}{j_{|m|n}'^{\,2}-m^2}\left[\frac{J_{|m|}\left(j_{|m|n}'\frac{r}{a}\right)}{J_{|m|}\left(j_{|m|n}'\right)}\right]^2\\
\label{NeumannDensities1_P}
\gamma_{nm}^N({\bf k})&\equiv\gamma_{nm}^N(k)=\frac{a^2}{\pi}\frac{j_{|m|n}'^{\,2}}{j_{|m|n}'^{\,2}-m^2}\left[\frac{ak}{j_{|m|n}'^{\,2}-(ak)^2}J_{|m|}'(ak)\right]^2;
\end{align}
\end{subequations}
in particular, similar to the 1D well \cite{Olendski1}, as $j_{01}'=0$ \cite{Abramowitz1}, the energy of the lowest state is just zero, 
\begin{align}\tag{46$'$}\label{eq:46'}
E_{10}^N&=0,\\
\intertext{and the position wave function is a constant,}
\tag{47a$'$}\label{eq:47a'}
\Psi_{10}^N(r,\varphi_{\bf r})&=\frac{1}{\pi^{1/2}a}.
\intertext{Due to this, a disturbance with such profile (TE$_{00}$-mode) can not propagate in electromagnetic waveguides \cite{Landau1}. As we shall see later, this orbital with its angle-independent momentum function}
\tag{47b$'$}\label{eq:47b'}
\Phi_{10}^N(k,\varphi_{\bf k})&=\frac{1}{\pi^{1/2}k}J_1(ak)
\end{align}
plays a special role in the quantum treatment too. Orthonormality from Equation~\eqref{Separation2_3} yields:
\begin{equation}\label{IntegralNeumann1}
\int_0^\infty\frac{\xi^3}{\left(j_{|m|n}'^{\,2}-\xi^2\right)\left(j_{|m|n'}'^{\,2}-\xi^2\right)}J_{|m|}'^{\,2}(\xi)d\xi=\frac{1}{2}\left[1-\left(\frac{m}{j_{|m|n}'}\right)^2\right]\delta_{nn'}.
\end{equation}
\begin{figure}
\centering
\includegraphics[width=\columnwidth]{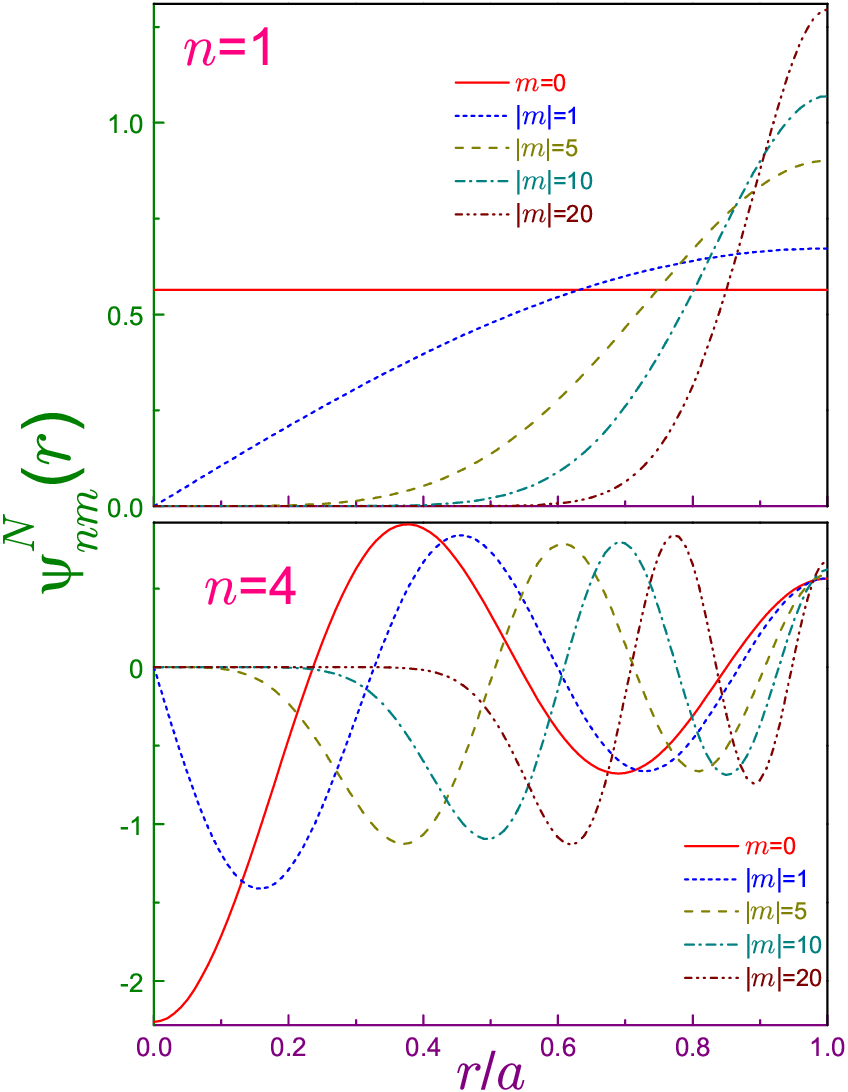}
\caption{\label{Fig_FunctionsNeumannR1}
Neumann position functions $\psi_{nm}^N(r)$ from Equation~\eqref{DirichletWaveForm2} in terms of dimensionless distance $r/a$. The same convention as in Figure~\ref{Fig_FunctionsDirichletR1} is adopted except that dash-dot-dotted curves are for $|m|=20$.}
\end{figure}

Similar to the Dirichlet case, rotationally symmetric family of states are the only ones that permit the particle to stay at the centre:
\begin{equation}\label{Equation1R_Neumann}
\rho_{nm}^N(0)=
\frac{1}{\pi a^2}\frac{1}{J_0^2\!\left(j_{0n}'\right)}\delta_{m0}.
\end{equation}
The density $\rho_{nm}^N(r)$ reaches its tops
\begin{subequations}\label{ExtremaNeumann1}
\begin{align}\label{ExtremaNeumann1_Magn}
\rho_{nmn'_{max}}^N&=\frac{1}{\pi a^2}\frac{j_{|m|n}'^{\,2}}{j_{|m|n}'^{\,2}-m^2}\left[\frac{J_{|m|}(j_{|m|n'}')}{J_{|m|}(j_{|m|n}')}\right]^2
\intertext{at}\label{ExtremaNeumann1_Pos}
r_{nmn'_{max}}^N&=\frac{j_{|m|n'}'}{j_{|m|n}'}\,a,
\end{align}
\end{subequations}
where the first peak, $n'=1$, is again the largest one. Considering the fundamental band, $n=1$, one derives \cite{Abramowitz1} that, at the huge magnetic index, the only maximum, which is always located at the rim, 
\begin{subequations}\label{ExtremaNeumann2}
\begin{align}\label{ExtremaNeumann2_Pos}
r_{1m1_{max}}^N&=a\\
\intertext{increases with $|m|$ as}
\label{ExtremaNeumann2_Magn}
\rho_{1m1_{max}}^N&\rightarrow\frac{0.6183}{\pi a^2}|m|^{2/3},\quad|m|\rightarrow\infty,
\end{align}
\end{subequations}
what echoes the Dirichlet geometry, Equation~\eqref{ExtremaDirichlet2_Magn}. Qualitatively, similar behavior is characteristic for the other bands too, as Figure~\ref{Fig_FunctionsNeumannR1} depicts on the examples of several position functions $\psi_{nm}^N(r)$. Straight horizontal line $\pi^{-1/2}=0.5641\ldots$ in the upper window portrays the zero-energy waveform, Eq.~\eqref{eq:47a'}, whose radial derivative vanishes not only at the rim, as it is required by the Neumann BC, but inside the disc too. It is important to underline here that the Neumann position functions are \textit{not} defined \textit{outside} the circle; in particular, they are \textit{not} zeros at $r>a$ as, if it were the case, the discontinuity at the interface will result in the infinite radial momentum whose operator was introduced in subsection~\ref{sec_WaveFunctionsDirichlet}.

\begin{figure}
\centering
\includegraphics[width=\columnwidth]{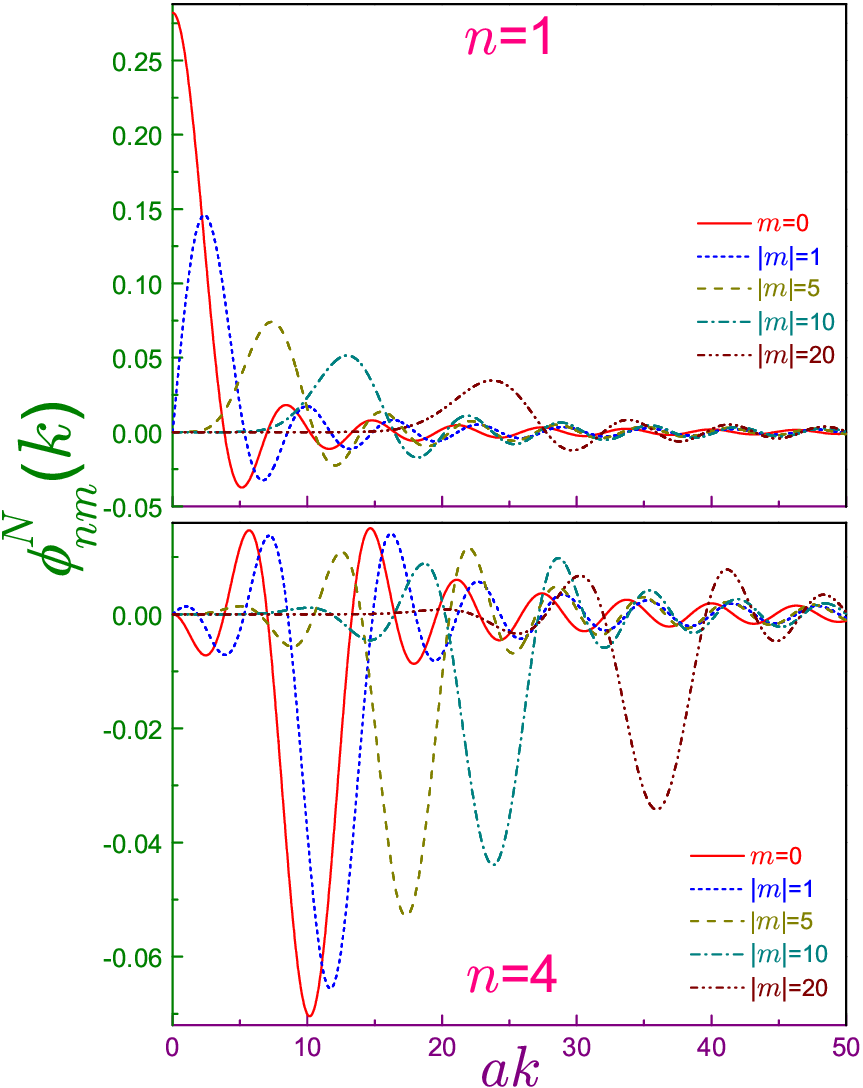}
\caption{\label{Fig_FunctionsNeumannP1}
Neumann momentum functions $\phi_{nm}^N(k)$ from Equation~\eqref{MomentumDirichletFunctions2} in terms of dimensionless wave vector $ak$. The same convention as in Figure~\ref{Fig_FunctionsNeumannR1} is adopted.}
\end{figure}

Compared to its Dirichlet counterpart, Equation~\eqref{AsymptoteDirichlet1}, momentum Neumann density at the infinity decays much slower:
\begin{equation}\label{AsymptoteNeumann1}
\gamma_{nm}^N(k)\sim\frac{1}{k^3},\quad k\rightarrow\infty,
\end{equation}
what has drastic consequences for the different behavior of the corresponding R\'{e}nyi and Tsallis entropies, Section~\ref{sec_Renyi}. Another distinction is the fact that the $n=1$, $m=0$ level is the only one that allows the particle to have zero momentum:
\begin{equation}\label{Equation1G_Neumann}
\gamma_{nm}^N(0)=\frac{a^2}{4\pi}\delta_{n1}\delta_{m0}.
\end{equation}

Among the infinite number of peaks of the Neumann momentum density $\gamma_{nm}^N(k)$, the $n$-th extremum, similar to the Dirichlet dot, is the largest one with its location $k_{nm_{max}}^N$ and magnitude $\gamma_{nm_{max}}^N$ for the huge $|m|$ and/or $n$ being, respectively,
\begin{subequations}\label{NeumannMomentumMax1}
\begin{align}\label{NeumannMomentumMax1_Location}
k_{nm_{max}}^N&=\frac{j_{|m|n}'}{a}\left[1+\frac{3m^2}{j_{|m|n}'^{\,4}-2m^2j_{|m|n}'^{\,2}+m^2(m^2+5)}\right],\quad |m|\,\,{\rm and/or}\,\,n\rightarrow\infty,\\
\gamma_{nm_{max}}^N&=\frac{a^2}{4\pi}\left[1-\left(\frac{m}{j_{|m|n}'}\right)^2\right]J_{|m|}^2\!\left(j_{|m|n}'\right)\nonumber\\
\label{NeumannMomentumMax1_Magnitude}
&\times\left(1\!+\!\frac{6m^4}{\left(j_{|m|n}'^{\,2}-m^2\right)\left[j_{|m|n}'^{\,4}-2m^2j_{|m|n}'^{\,2}+m^2(m^2+5)\right]}\!\right),\,|m|\,\,{\rm and/or}\,\,n\rightarrow\infty.
\end{align}
\end{subequations}
In the limit $|m|\rightarrow\infty$, the product $\left[1-\left(m/j_{|m|n}'\right)^2\right]J_{|m|}^2\!\left(j_{|m|n}'\right)$ fades to zero as $|m|^{-4/3}$ \cite{Abramowitz1}, which is the same rate as that of the leading Dirichlet term in Equation~\eqref{DirichletMomentumMax1_Magnitude}. The decrease of the global extremum and its shift to the faster momenta with the azimuthal index increasing are exemplified in Figure~\ref{Fig_FunctionsNeumannP1}. Observe that due to the slow decrease of the Neumann wave vector density, Equation~\eqref{AsymptoteNeumann1}, the radial integral $\int_0^\infty k^3\gamma_{nm}^N(k)dk$ in the corresponding second-order moment $\left\langle k^2\right\rangle$ does diverge. Accordingly, the Heisenberg uncertainty, Equation~\eqref{Heisenberg1}, for this BC looses its meaning for all states if the momentum expectation values are calculated in the wave vector representation. Contrary, the corresponding Shannon, Equation~\eqref{ShannonInequality1}, R\'{e}nyi, Equation~\eqref{RenyiUncertainty1}, and Tsallis, Equation~\eqref{TsallisInequality1}, inequalities do make sense, as shown below. Let us also point out that the averaging of the linear radial momentum operator, Equation~\eqref{MomentumOperator1_1}, over the position waveform, Equation~\eqref{eq:47a'}, of  the zero-energy Neumann level does not vanish, $\left\langle \widehat{k}_r\right\rangle_{1,0}^N=-i/a$, what adds more arguments into the discussion of its physical meaning. This consideration and that from section~\ref{sec_WaveFunctionsDirichlet} show that the number of inconsistencies and controversies of the Heisenberg relation and the radial momentum definition gets larger for the 2D geometry as compared to the 1D well \cite{Bialynicki3,Bialynicki4,Olendski4}. All these problems are eliminated if one employs quantum-information measures $S$, $R$ and $T$.

\section{Shannon entropy, Fisher information and Onicescu energy}\label{sec_Shannon}
\begin{sidewaystable}
\caption{Position $S_{\rho_{nm}}^D$ and momentum $S_{\gamma_{nm}}^D$ entropies together with their sum $S_{t_{nm}}^D$ for the unit-radius Dirichlet disc}
\centering 
\begin{tabular}{|c||c|c|c||c|c|c||c|c|c||c|c|c|}
\hline 
\multirow{3}{1em}{$|m|$}& \multicolumn{12}{c|}{Principal quantum number $n$}\\ 
\cline{2-13} 
 &\multicolumn{3}{|c||}{$n=1$}&\multicolumn{3}{c||}{$n=2$}&\multicolumn{3}{c||}{$n=3$}&\multicolumn{3}{c|}{$n=4$}\\ 
\cline{2-13}
&$S_\rho^D$&$S_\gamma^D$&$S_t^D$&$S_\rho^D$&$S_\gamma^D$&$S_t^D$&$S_\rho^D$&$S_\gamma^D$&$S_t^D$&$S_\rho^D$&$S_\gamma^D$&$S_t^D$\\
\hline 
0&0.5942&3.8232&4.4174&0.5589&5.1255&5.6844&0.5488&5.7198&6.2686&0.5441&6.0988&6.6429\\
1&0.8103&4.6258&5.4361&0.6971&5.4655&6.1626&0.6494&5.9292&6.5786&0.6230&6.2487&6.8717\\
2&0.8215&5.0623&5.8838&0.7359&5.7100&6.4459&0.6888& 6.0972&6.7860&0.6592&6.3760&7.0352\\
3&0.7934&5.3710&6.1644&0.7418&5.9050&6.6468&0.7037&6.2395&6.9432&0.6768&6.4878&7.1646\\
4&0.7547&5.6124&6.3671&0.7336&6.0685&6.8021&0.7063&6.3637&7.0700&0.6840&6.5879&7.2719\\
5&0.7136&5.8116&6.5252&0.7185&6.2100&6.9285&0.7019&6.4743&7.1762&0.6849&6.6788&7.3637\\
10&0.5252&6.4906&7.0158&0.6157&6.7300&7.3457&0.6419&6.9010&7.5429&0.6498&7.0420&7.6918\\
20&0.2560&7.2449&7.5009&0.4252&7.3605&7.7857&0.4986&7.4507&7.9493&0.5386&7.5320&8.0706\\
30&0.0685&7.7131&7.7816&0.2763&7.7725&8.0488&0.3753&7.8235&8.1988&0.4343&7.8743&8.3086\\
\hline
\end{tabular}
\label{Table_EntropyDirichlet}
\end{sidewaystable}

Before discussing 2D measures, for comparative reasons it does make sense to recall what are their 1D counterparts \cite{Olendski4,Olendski5}. A noteworthy property is the same Shannon position entropy of all Dirichlet orbitals and any excited Neumann level of the 1D well of the width $a$:
\begin{subequations}\label{1D}
\begin{align}\label{1D_ShannonDirichletX1}
{\rm 1D:}&\quad S_{\rho_n}^D=S_{\rho_{n+1}}^N=\ln a-1+\ln2,
\intertext{whereas the ground-state Neumann functional reads:}
\label{1D_ShannonNeumannX1}
{\rm 1D:}&\quad S_{\rho_1}^N=\ln a.
\intertext{The sum $S_{t_n}$ increases with $n$ and the correspoding uncertainty relation, Equation~\eqref{ShannonInequality1}, is always satisfied. Position Fisher informations depend quadratically on the principal index}
{\rm 1D:}&\quad I_{\rho_n}^D=\frac{4n^2\pi^2}{a^2},\quad I_{\rho_n}^N=\frac{4(n-1)^2\pi^2}{a^2},
\intertext{what means that they are essentially proportional to the associated energies:}
\label{1D_Energies1}
{\rm 1D:}&\quad E_n^D=\frac{\pi^2\hbar^2n^2}{2m^*a^2},\quad E_n^N=\frac{\pi^2\hbar^2(n-1)^2}{2m^*a^2}.
\intertext{Dirichlet Fisher momentum component is an increasing function of $n$ whereas the Neumann one has its largest magnitude at the first excited level monotonically decreasing for the higher lying orbitals in such a way that both Dirichlet and Neumann products $I_{\rho_n}I_{\gamma_n}$ grow at any quantum number. Similar to the Shannon entropy, 1D position Onicescu energy is the same for all Dirichlet and excited Neumann states:}
{\rm 1D:}&\quad O_{\rho_n}^D=O_{\rho_{n+1}}^N=\frac{3}{2a},\quad O_{\rho_1}^N=\frac{1}{a},
\end{align}
\end{subequations}
with the product $O_{\rho_n}O_{\gamma_n}$ being a monotonically decreasing function of the quantum index for either BC.

\subsection{Dirichlet BC}\label{sec_ShannonDirichlet}
Expressions for the Shannon entropies read:
\begin{subequations}\label{DirichletShannon1}
\begin{eqnarray}\label{DirichletShannon1X}
S_{\rho_{nm}}^D&=&2\ln a+\ln\pi-2\!\!\int_0^1\!\!z\!\left[\frac{J_{|m|}(j_{|m|n}z)}{J_{|m|+1}(j_{|m|n})}\right]^2\!\ln\!\frac{J_{|m|}^2(j_{|m|n}z)}{J_{|m|+1}^2(j_{|m|n})}dz\\
S_{\gamma_{nm}}^D&=&-2\ln a+\ln\pi\nonumber\\
\label{DirichletShannon1K}
&-&2\!\!\int_0^\infty\!\!\xi\!\!\left[\frac{j_{|m|n}}{j_{|m|n}^2-\xi^2}J_{|m|}(\xi)\right]^2\!\!\ln\!\!\left(\!\left[\frac{j_{|m|n}}{j_{|m|n}^2-\xi^2}J_{|m|}(\xi)\!\right]^2\right)\!\!d\xi,
\end{eqnarray}
\end{subequations}
where $z=r/a$ and $\xi=ak$ are dimensionless distance and wave vector, respectively. These equations manifest that the whole dependence on the dot radius for both entropies is determined by the positive or negative double (as $d=2$) logarithm of $a$. Accordingly, as a result of the dependencies from Equation~\eqref{DirichletShannon1}, the sum of the two entropies $S_t$ is a dimensionless scaling-independent quantity, as it should be \cite{Dodonov1}.

In Table \ref{Table_EntropyDirichlet}, the values of the position $S_{\rho_{nm}}^D$ and momentum $S_{\gamma_{nm}}^D$ Shannon entropies together with their sum $S_{t_{nm}}^D$ are provided for $m$ from $0$ to $30$ and $n=1-4$ where the dot radius is assumed to be equal to unity. Since there are no in known literature analytic expressions for the corresponding integrals \cite{Abramowitz1,Gradshteyn1,Prudnikov2,Prudnikov3,Brychkov1}, a direct numerical quadrature was employed in computing them. Our results show that the momentum entropy for the one fixed quantum number ($m$ or $n$) is a monotonically increasing function of the second index. Since the logarithm argument in Equation~\eqref{DirichletShannon1K} is always smaller than unity, $S_{\gamma_{nm}}^D+2\ln a$ is greater than $\ln\pi=1.1447\ldots$ for any combination of $|m|$ and $n$. An increase of dimensionality wipes out an identity of the 1D position Shannon entropies from Equation~\eqref{1D_ShannonDirichletX1}: $S_{\rho_{nm}}$ exhibits at the fixed $n$ a nonmonotonic dependence on $|m|$; namely, its initial growth at the small and moderate magnetic quantum number turns into a decrease with the subsequent growth of $|m|$. For the fixed $|m|$, the position entropy generally decreases with the principal quantum index: for the small and moderate magnetic index this dependence is a monotonic one, whereas for the higher $|m|$ values, a maximum of $S_\rho$ is observed on the discrete $n$ axis. We will return to the explanation of these two behaviors at the end of this subsection. The sum of the two entropies is an increasing function of both indexes. Of course, inequality from Equation~\eqref{ShannonInequality1} with its right-hand side  $2(1+\ln\pi)=4.2894\ldots$ is always satisfied for this BC with the lowest level, $n=1$, $m=0$, coming closest to saturating it with $S_{t_{10}}^D=4.4174\ldots$. Let us underline that the left-hand side of the Shannon uncertainty relation stays finite for all orbitals, including azimuthally symmetric ones, for which the Heisenberg inequality does not have any meaning, as discussed in section~\ref{sec_WaveFunctionsDirichlet}.

\begin{sidewaystable}
\caption{Position $I_{\rho_{nm}}^D$ and momentum $I_{\gamma_{nm}}^D$ Fisher informations together with their product for the unit-radius Dirichlet disc}
\centering 
\begin{tabular}{|c||c|c|c||c|c|c||c|c|c||c|c|c|}
\hline 
\multirow{3}{1em}{$|m|$}& \multicolumn{12}{c|}{Principal quantum number $n$}\\ 
\cline{2-13} 
 &\multicolumn{3}{|c||}{$n=1$}&\multicolumn{3}{c||}{$n=2$}&\multicolumn{3}{c||}{$n=3$}&\multicolumn{3}{c|}{$n=4$}\\ 
\cline{2-13}
&$I_\rho^D$&$I_\gamma^D$&$I_\rho^DI_\gamma^D$&$I_\rho^D$&$I_\gamma^D$&$I_\rho^DI_\gamma^D$&$I_\rho^D$&$I_\gamma^D$&$I_\rho^DI_\gamma^D$&$I_\rho^D$&$I_\gamma^D$&$I_\rho^DI_\gamma^D$\\
\hline 
0&0.2313E+2&0.8722&0.2018E+2&0.1219E+3&1.2458&0.1518E+3&0.2995E+3&1.2977&0.3887E+3&0.5562E+3&1.3142&0.7309E+3\\
1&0.3807E+2&0.7884&0.3002E+2&0.1565E+3&1.1708&0.1832E+3&0.3538E+3&1.2560&0.4444E+3&0.6302E+3&1.2883&0.8119E+3\\
2&0.5337E+2&0.7267&0.3879E+2&0.1912E+3&1.1075&0.2118E+3&0.4083E+3&1.2148&0.4960E+3&0.7043E+3&1.2603&0.8876E+3\\
3&0.6912E+2&0.6782&0.4688E+2&0.2263E+3&1.0535&0.2384E+3&0.4629E+3&1.1759&0.5443E+3&0.7786E+3&1.2320&0.9592E+3\\
4&0.8533E+2&0.6387&0.5450E+2&0.2618E+3&1.0066&0.2635E+3&0.5178E+3&1.1397&0.5902E+3&0.8531E+3&1.2044&0.1027E+4\\
5&0.1020E+3&0.6055&0.6174E+2&0.2977E+3&0.9655&0.2874E+3&0.5731E+3&1.1062&0.6340E+3&0.9279E+3&1.1779&0.1093E+4\\
10&0.1910E+3&0.4934&0.9426E+2&0.4838E+3&0.8154&0.3945E+3&0.8559E+3&0.9712&0.8313E+3&0.1308E+4&1.0629&0.1390E+4\\
20&0.3928E+3&0.3798&0.1492E+3&0.8871E+3&0.6471&0.5741E+3&0.1454E+4&0.8001&0.1164E+4&0.2100E+4&0.9016&0.1893E+4\\
30&0.6189E+3&0.3183&0.1970E+3&0.1326E+4&0.5500&0.7291E+3&0.2093E+4&0.6931&0.1451E+4&0.2934E+4&0.7937&0.2329E+4\\
\hline
\end{tabular}
\label{Table_FisherDirichlet}
\end{sidewaystable}

Position Fisher informations \cite{TorresArenas1}
\begin{subequations}\label{FisherDirichlet1}
\begin{align}\label{FisherDirichlet1_1}
I_{\rho_{nm}}^D&=\frac{8}{a^2}\frac{j_{|m|n}^2}{J_{|m|+1}^2(j_{|m|n})}\int_0^1zJ_{|m|}'^{\,2}(j_{|m|n}z)dz
\intertext{can be calculated analytically as:}
\label{FisherDirichlet1_2}
I_{\rho_{nm}}^D&=\frac{4}{a^2}\left(j_{|m|n}^2-\frac{|m|}{J_{|m|}'^{\,2}(j_{|m|n})}\left[1-J_0^2(j_{|m|n})-2\sum_{k=1}^{|m|-1}J_k^2(j_{|m|n})\right]\right).
\end{align}
\end{subequations}
This expression, in the wake of the discussion of Equations~\eqref{MomentumOperator2}-\eqref{MomentumOperator3}, straightforwardly points at the direct correlation between $I_{\rho_{nm}}^D$, which is inversely proportional to the square of the dot radius, and the associated total energy $E_{nm}^D$, Equation~\eqref{DirichletSpectrum1}, less its azimuthal component. Numerical results from Equation~\eqref{FisherDirichlet1_2} together with their momentum counterparts that are proportional to $a^2$ are given in Table~\ref{Table_FisherDirichlet} where also the radius-independent products of the two informations are provided. Obviously, $I_{\rho_{nm}}^D$ grows both with $|m|$ and $n$ whereas the momentum component is an increasing (decreasing) function of the principal (magnetic) index in such a way that the product $I_\rho I_\gamma$ gets greater for either of the quantum numbers enlarging. The decrease of the momentum component with $|m|$ growing can be understood from Figure~\ref{Fig_FunctionsDirichletP1} and Equation~\eqref{DirichletMomentumMax1_Magnitude} that show a suppression of oscillations for the larger $|m|$ and since, due to the presence of the gradients in Equations~\eqref{Fisher1}, Fisher information is a local measure of distribution that is very sensitive to speed of change of the corresponding density, $I_\gamma$ drops for the larger $|m|$.

\begin{sidewaystable}
\caption{Position $O_{\rho_{nm}}^D$ and momentum $O_{\gamma_{nm}}^D$ Onicescu energies together with their product for the unit-radius Dirichlet disc}
\centering 
\begin{tabular}{|c||c|c|c||c|c|c||c|c|c||c|c|c|}
\hline 
\multirow{3}{1em}{$|m|$}& \multicolumn{12}{c|}{Principal quantum number $n$}\\ 
\cline{2-13} 
 &\multicolumn{3}{|c||}{$n=1$}&\multicolumn{3}{c||}{$n=2$}&\multicolumn{3}{c||}{$n=3$}&\multicolumn{3}{c|}{$n=4$}\\ 
\cline{2-13}
&$O_\rho^D$&$O_\gamma^D$&$O_\rho^DO_\gamma^D$&$O_\rho^D$&$O_\gamma^D$&$O_\rho^DO_\gamma^D$&$O_\rho^D$&$O_\gamma^D$&$O_\rho^DO_\gamma^D$&$O_\rho^D$&$O_\gamma^D$&$O_\rho^DO_\gamma^D$\\
\hline 
0&0.6679&0.2909E-1&0.1943E-1&0.8730&0.7242E-2&0.6322E-2&0.9821&0.4178E-2&0.4103E-2&1.0567&0.2969E-2&0.3137E-2\\
1&0.4939&0.1188E-1&0.5868E-2&0.6245&0.5263E-2&0.3287E-2&0.7092&0.3461E-2&0.2454E-2&0.7718&0.2596E-2&0.2004E-2\\
2&0.4871&0.7649E-2&0.3726E-2&0.5777&0.4197E-2&0.2424E-2&0.6449&0.2970E-2&0.1915E-2&0.6978&0.2312E-2&0.1613E-2\\
3&0.5023&0.5631E-2&0.2829E-2&0.5654&0.3497E-2&0.1977E-2&0.6192&0.2603E-2&0.1612E-2&0.6642&0.2085E-2&0.1384E-2\\
4&0.5237&0.4438E-2&0.2324E-2&0.5655&0.2996E-2&0.1694E-2&0.6086&0.2318E-2&0.1411E-2&0.6468&0.1898E-2&0.1228E-2\\
5&0.5471&0.3647E-2&0.1995E-2&0.5714&0.2619E-2&0.1496E-2&0.6055&0.2088E-2&0.1264E-2&0.6380&0.1742E-2&0.1111E-2\\
10&0.6661&0.1869E-2&0.1245E-2&0.6279&0.1588E-2&0.9970E-3&0.6299&0.1387E-2&0.8740E-3&0.6414&0.1231E-2&0.7893E-3\\
20&0.8784&0.8878E-3&0.7798E-3&0.7576&0.8582E-3&0.6502E-3&0.7191&0.8125E-3&0.5843E-3&0.7044&0.7639E-3&0.5381E-3\\
30&1.0633&0.5585E-3&0.5939E-3&0.8788&0.5724E-3&0.5030E-3&0.8106&0.5634E-3&0.4569E-3&0.7587&0.5255E-3&0.3986E-3\\
\hline
\end{tabular}
\label{Table_OnicescuDirichlet}
\end{sidewaystable}

A transition to higher dimensionality lifts the independence of the position Onicescu energy on the orbital, which for the dot reads:
\begin{equation}\label{OnicescuDirichlet1}
O_{\rho_{nm}}^D=\frac{1}{a^2}\frac{2}{\pi J_{|m|}'^{\,4}(j_{|m|n})}\int_0^1zJ_{|m|}^4(j_{|m|n}z)dz.
\end{equation}
Most characteristic features of the Onicescu energies presented in Table~\ref{Table_OnicescuDirichlet} are:
\begin{itemize}
\item{both $O_\gamma^D$, which are proportional to $a^2$, and radius-independent products $O_\rho^DO_\gamma^D$ decrease with $|m|$ and $n$ growing;}
\item{$O_\rho^D$ is a concave function of the magnetic quantum number with its minimum (whose magnitude increases with $n$) being achieved at the larger $|m|$ for the greater principal index;}
\item{for small and moderate $|m|$, the position disequilibrium monotonically increases with $n$; however, for the greater magnetic index (for example, $|m|=10$ and above) the Onicescu energy $O_\rho^D$ has a minimum on the discrete $n$ axis. This behavior is opposite to the one discussed for the Shannon entropy.}
\end{itemize}
The first dependence can be understood by recalling that, physically, the Onicescu energy describes a deviation from the equilibrium or the most probable distribution, which is a uniform one. Figure~\ref{Fig_FunctionsDirichletP1} shows that for the larger $|m|$ the structure of the momentum waveforms gets more uniform thus decreasing $O_\gamma$. The second item is treated in a similar way; say, for $n=1$ (upper panel in Figure~\ref{Fig_FunctionsDirichletR1}) the wave function for $|m|=1$ (dashed line) is much more uniform than its cylindrically symmetric counterpart (solid curve): the former one, which is almost symmetric with respect to $r=a/2$, vanishes at the ends of the interval with its maximum being smaller than the extremum for $m=0$ dependence that exhibits a continuous decrease. As a result, $O_{\rho_{11}}^D<O_{\rho_{10}}^D$. The unlimited increase of the position Onicescu energy (or unrestricted decrease of the Shannon entropy) at the large magnetic index is explained by the mentioned above accumulation of the density $\rho$ near the edge: one sees that for $|m|\gg1$ the function $\Psi$ is almost zero along the $r$ axis and only near the interface it has a sharp maximum which destroys the uniformity thus rising $O_\rho$ or precipitously sinking $S_\rho$. In the extreme limit of the magnetic index tending to infinity, one has:
\begin{equation}\label{ExtremeLimit1}
\left\{
\begin{array}{c}
\psi_{nm}(r)\\
\phi_{nm}(k)\\
S_{\rho_{nm}}\\
S_{\gamma_{nm}}\\
O_{\rho_{nm}}\\
O_{\gamma_{nm}}
\end{array}
\right\}
\xrightarrow[]{|m|\rightarrow\infty}
\left\{
\begin{array}{c}
\frac{1}{2^{1/2}}\delta\left(\frac{r}{a}-1\right)\\
0\\
-\infty\\
\infty\\
\infty\\
0
\end{array}
\right\}.
\end{equation}
Here, $\delta(z)$ is a Dirac $\delta$-function. Accumulation of the position dependence at the edge and fading of the momentum waveform were analytically discussed in subsection~\ref{sec_WaveFunctionsDirichlet}. Physically, Shannon entropy describes quantitatively the lack of our knowledge about the phenomenon; so, its extremely large negative (positive) values for the position (momentum) component manifest that we know precisely where the particle is located (know nothing about its momentum). The deviation of the position and momentum Shannon entropies in the opposite directions in this limit leads to the increase of their sum, as exemplified by Table~\ref{Table_EntropyDirichlet}. In general, the growth of the principal or magnetic quantum number means a shift to the higher energies, as it follows from Equation~\eqref{DirichletSpectrum1} and properties of the zeros of Bessel functions \cite{Abramowitz1}. Accordingly, with $|m|$ or $n$ tending to infinity, one departs more and more from the quantum regime approaching the classical behavior when the position of the particle can be exactly specified what corresponds to the infinitely negative position entropy.

\begin{sidewaystable}
\caption{Position $S_{\rho_{nm}}$ and momentum $S_{\gamma_{nm}}$ entropies together with their sum $S_{t_{nm}}$ for the unit-radius Neumann disc}
\centering 
\begin{tabular}{|c||c|c|c||c|c|c||c|c|c||c|c|c|}
\hline 
\multirow{3}{1em}{$|m|$}&\multicolumn{12}{c|}{Principal quantum number $n$}\\ 
\cline{2-13} 
 &\multicolumn{3}{|c||}{$n=1$} & \multicolumn{3}{c||}{$n=2$}& \multicolumn{3}{c||}{$n=3$}& \multicolumn{3}{c|}{$n=4$}\\ 
\cline{2-13}
&$S_\rho^N$&$S_\gamma^N$&$S_t^N$&$S_\rho^N$&$S_\gamma^N$&$S_t^N$&$S_\rho^N$&$S_\gamma^N$&$S_t^N$&$S_\rho^N$&$S_\gamma^N$&$S_t^N$\\
\hline 
0&1.1447&4.2880&5.4327&0.5647&5.3874&5.9521&0.5509&5.8924&6.4433&0.5452&6.2274&6.7726\\
1&1.0379&5.2247&6.2626&0.7411&5.6814&6.4225&0.6693&6.0761&6.7454&0.6344&6.3613&6.9957\\
2&0.9092&5.7280&6.6372&0.7724&5.9167&6.6891&0.7094&6.2344&6.9438&0.6725&6.4810&7.1535\\
3&0.7992&6.0786&6.8778&0.7659&6.1113&6.8772&0.7210&6.3724&7.0934&0.6891&6.5885&7.2776\\
4&0.7055&6.3489&7.0544&0.7456&6.2772&7.0228&0.7191&6.4949&7.2140&0.6944&6.6862&7.3806\\
5&0.6241&6.5702&7.1943&0.7199&6.4221&7.1420&0.7101&6.6053&7.3154&0.6930&6.7759&7.4689\\
10&0.3285&7.3085&7.6370&0.5806&6.9566&7.5372&0.6310&7.0355&7.6665&0.6465&7.1385&7.7850\\
20&-0.0203&8.1106&8.0903&0.3550&7.6054&7.9604&0.4659&7.5919&8.0578&0.5203&7.6318&8.1521\\
\hline
\end{tabular}
\label{Table_EntropyNeumann}
\end{sidewaystable}

\subsection{Neumann BC}\label{sec_ShannonNeumann}
Due to the simplicity of its wave function, Equation~\eqref{eq:47a'}, ground-state position functionals can be elementarily calculated:
\begin{subequations}\label{NeumannAnalytic1}
\begin{align}\label{NeumannAnalytic1_1}
S_{\rho_{10}}^N&=2\ln a+\ln\pi\\
\label{NeumannAnalytic1_2}
I_{\rho_{10}}^N&=0\\
\label{NeumannAnalytic1_3}
O_{\rho_{10}}^N&=\frac{1}{\pi a^2}
\intertext{together with the momentum Fisher information}
\label{NeumannAnalytic1_4}
I_{\gamma_{10}}^N&=2a^2.
\end{align}
\end{subequations}
Table \ref{Table_EntropyNeumann} manifests that the sum of the Shannon entropies for all states does obey the corresponding inequality from Equation~\eqref{ShannonInequality1}, as expected. Ground-orbital Neumann sum of $5.4327$ is greater than its Dirichlet counterpart from Table~\ref{Table_EntropyDirichlet}. This rule holds true for any other level and is valid for the momentum components too. Position entropy of any state with $n\geq2$ is, similar to the Dirichlet BC, a nonmonotonic function of the magnetic index, whereas for $n=1$ orbitals it steadily decreases with $|m|$ and turns negative at $|m|=20$ when $S_{\rho_{nm}}^D$ is still positive. Negative values of the position component mean that the parts of the function $|\psi_{1,20}^N(r)|$ from Equation~\eqref{DirichletWaveForm2}, which are greater than unity [and which, in the case of the lowest Neumann band, accumulate near the surface, see upper window of Figire~\ref{Fig_FunctionsNeumannR1}], overweigh in their contribution to the integral from Equation~\eqref{Shannon1_R} those regions where $|\psi_{1,20}^N(r)|<1$. For other principal quantum numbers, $n\geq2$, this takes place at much greater magnetic indexes that, accordingly, are not shown in Tables~\ref{Table_EntropyDirichlet} and \ref{Table_EntropyNeumann}. Let us also point out that at small and moderate $|m|$ the Neumann momentum component $S_{\gamma_{nm}}$  and the sum $S_{t_{nm}}$ monotonically increase with $n$, as they always do for the Dirichlet BC, but at the higher  magnetic number, both become concave functions of the band. The difference between the Dirichlet and Neumann data decreases for the growing quantum numbers $n$  and $|m|$ what is explained by the smaller sensitivity to the BC of the higher lying levels.

\begin{sidewaystable}
\caption{Position $I_{\rho_{nm}}$ and momentum $I_{\gamma_{nm}}$ Fisher informations together with their product for the unit-radius Neumann dot}
\centering 
\begin{tabular}{|c||c|c|c||c|c|c||c|c|c||c|c|c|}
\hline 
\multirow{3}{1em}{$|m|$}& \multicolumn{12}{c|}{Principal quantum number $n$}\\ 
\cline{2-13} 
 &\multicolumn{3}{|c||}{$n=1$}&\multicolumn{3}{c||}{$n=2$}&\multicolumn{3}{c||}{$n=3$}&\multicolumn{3}{c|}{$n=4$}\\ 
\cline{2-13}
&$I_\rho^N$&$I_\gamma^N$&$I_\rho^NI_\gamma^N$&$I_\rho^N$&$I_\gamma^N$&$I_\rho^NI_\gamma^N$&$I_\rho^N$&$I_\gamma^N$&$I_\rho^NI_\gamma^N$&$I_\rho^N$&$I_\gamma^N$&$I_\rho^NI_\gamma^N$\\
\hline 
0&0&2&0&0.5873E+2&1.3333&0.7830E+2&0.1969E+3&1.3333&0.2625E+3&0.4140E+3&1.3333&0.5520E+3\\
1&4.1493&1.7629&7.3146&0.8338E+2&1.2950&0.1080E+3&0.2413E+3&1.3163&0.3176E+3&0.4782E+3&1.3240&0.6331E+3\\
2&8.7248&1.6200&0.1413E+2&0.1082E+3&1.2379&0.1339E+3&0.2858E+3&1.2842&0.3670E+3&0.5424E+3&1.3040&0.7073E+3\\
3&0.1370E+2&1.5180&0.2079E+2&0.1333E+3&1.1837&0.1578E+3&0.3305E+3&1.2487&0.4127E+3&0.6068E+3&1.2800&0.7766E+3\\
4&0.1901E+2&1.4393&0.2737E+2&0.1589E+3&1.1352&0.1804E+3&0.3756E+3&1.2135&0.4558E+3&0.6714E+3&1.2545&0.8423E+3\\
5&0.2463E+2&1.3754&0.3388E+2&0.1849E+3&1.0919&0.2019E+3&0.4211E+3&1.1799&0.4968E+3&0.7364E+3&1.2290&0.9050E+3\\
10&0.5634E+2&1.1698&0.6591E+2&0.3212E+3&0.9330&0.2997E+3&0.6547E+3&1.0411&0.6817E+3&0.1067E+4&1.1134&0.1188E+4\\
20&0.1328E+3&0.9692&0.1287E+3&0.6221E+3&0.7566&0.4707E+3&0.1154E+4&0.8640&0.9972E+3&0.1761E+4&0.9483&0.1670E+4\\
\hline
\end{tabular}
\label{Table_FisherNeumann}
\end{sidewaystable}
Table~\ref{Table_FisherNeumann} that presents Fisher informations for the Neumann disc shows, in addition to the earlier obtained analytic results, Equations~\eqref{NeumannAnalytic1_2} and \eqref{NeumannAnalytic1_4}, a simple expression for the momentum components of the excited states with $m=0$:
\begin{equation}\label{NeumannAnalytic1_5}
I_{\gamma_{n0}}^N=\frac{4}{3}a^2,\quad n\geq2.
\end{equation}
All other features are qualitatively similar to those described in section~\ref{sec_ShannonDirichlet} for the Dirichlet dot.

\begin{sidewaystable}
\caption{Position $O_{\rho_{nm}}$ and momentum $O_{\gamma_{nm}}$ Onicescu energies together with their product for the unit-radius Neumann disc}
\centering 
\begin{tabular}{|c||c|c|c||c|c|c||c|c|c||c|c|c|}
\hline 
\multirow{3}{1em}{$|m|$}& \multicolumn{12}{c|}{Principal quantum number $n$}\\ 
\cline{2-13} 
 &\multicolumn{3}{|c||}{$n=1$}&\multicolumn{3}{c||}{$n=2$}&\multicolumn{3}{c||}{$n=3$}&\multicolumn{3}{c|}{$n=4$}\\ 
\cline{2-13}
&$O_\rho^N$&$O_\gamma^N$&$O_\rho^NO_\gamma^N$&$O_\rho^N$&$O_\gamma^N$&$O_\rho^NO_\gamma^N$&$O_\rho^N$&$O_\gamma^N$&$O_\rho^NO_\gamma^N$&$O_\rho^N$&$O_\gamma^N$&$O_\rho^NO_\gamma^N$\\
\hline 
0&0.3183&0.3658E-1&0.1164E-1&0.8122&0.8427E-2&0.6844E-2&0.9429&0.4746E-2&0.4474E-2&1.0275&0.3296E-2&0.3387E-2\\
1&0.3718&0.1280E-1&0.4759E-2&0.5725&0.6057E-2&0.3468E-2&0.6735&0.3889E-2&0.2619E-2&0.7446&0.2859E-2&0.2129E-2\\
2&0.4356&0.7629E-2&0.3323E-2&0.5397&0.4709E-2&0.2541E-2&0.6155&0.3291E-2&0.2026E-2&0.6741&0.2523E-2&0.1701E-2\\
3&0.4946&0.5354E-2&0.2648E-2&0.5382&0.3840E-2&0.2067E-2&0.5953&0.2850E-2&0.1697E-2&0.6438&0.2257E-2&0.1453E-2\\
4&0.5492&0.4078E-2&0.2240E-2&0.5470&0.3233E-2&0.1768E-2&0.5893&0.2511E-2&0.1480E-2&0.6294&0.2041E-2&0.1284E-2\\
5&0.6006&0.3268E-2&0.1963E-2&0.5602&0.2785E-2&0.1560E-2&0.5901&0.2242E-2&0.1323E-2&0.6230&0.1862E-2&0.1160E-2\\
10&0.8238&0.1561E-2&0.1286E-2&0.6428&0.1614E-2&0.1037E-2&0.6289&0.1445E-2&0.9089E-3&0.6361&0.1286E-2&0.8181E-3\\
20&1.1846&0.7014E-3&0.8309E-3&0.8066&0.8373E-3&0.6754E-3&0.7364&0.8215E-3&0.6049E-3&0.7113&0.7801E-3&0.5549E-3\\
\hline
\end{tabular}
\label{Table_OnicescuNeumann}
\end{sidewaystable}

Finally, Onicescu energies for the Neumann BC are given in Table~\ref{Table_OnicescuNeumann}. Qualitatively, their properties resemble their Dirichlet counterparts but quantitatively modified by the different type of the edge requirement; for example, position disequilibrium becomes a concave function of the principal index already at $|m|=4$. As was the case with the other quantum information measures, the difference between the two types of the BCs diminishes for the larger indexes $|m|$ and $n$. As a final observation, let us state that the limit of the extremely huge magnetic indexes from Equation~\eqref{ExtremeLimit1} remains true for the Neumann BC too.

\begin{figure}
\centering
\includegraphics[width=\columnwidth]{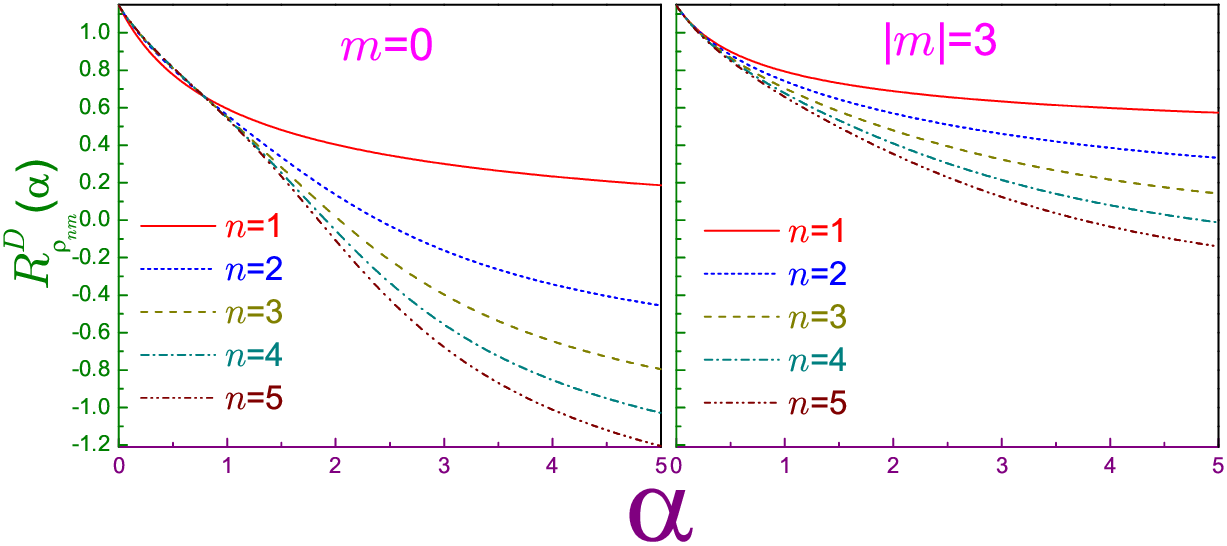}
\caption{\label{Fig_RenyiPositionDirichlet1}R\'{e}nyi position entropy of the Dirichlet disc $R_{\rho_{nm}}^D(\alpha)$ as a function of the parameter $\alpha$ for the states with $m=0$ (left panel) and $|m|=3$ (right window) where the solid lines are for $n=1$ levels, the dotted curves depict $n=2$ orbitals, the dashed dependencies exhibit $n=3$ states, the dash-dotted lines are for the $n=4$ levels,  and the dash-dot-dotted ones - for $n=5$. The dot radius is assumed to be equal to unity, $a\equiv1$.}
\end{figure}

\section{R\'{e}nyi and Tsallis entropies}\label{sec_Renyi}
\subsection{Dirichlet dot}\label{sec_RenyiDirichlet}
Contrary to the 1D well \cite{Olendski1}, the dependence of the R\'{e}nyi position entropy of the Dirichlet disc
\begin{equation}\label{DirichletRenyiPosition1}
R_{\rho_{nm}}^D(\alpha)=2\ln a+\ln\pi+\frac{1}{1-\alpha}\ln\!\left(\!2\!\int_0^1z\!\left[\frac{J_{|m|}^2(j_{|m|n}z)}{J_{|m|+1}^2(j_{|m|n})}\right]^\alpha\!\!dz\right)
\end{equation}
is a distinct function of each corresponding level what is exemplified in  Figure~\ref{Fig_RenyiPositionDirichlet1}. As it follows from the most general properties \cite{Renyi2}, $R_{\rho_{nm}}^D(\alpha)$ monotonically decreases with its argument increasing. At the vanishing coefficient, it approaches orbital-independent $\alpha=0$ limit as
\begin{equation}\label{DirichletRenyiPositionAsymptote0}
R_{\rho_{nm}}^D(\alpha\rightarrow0)=2\ln a+\ln\pi+2\alpha\int_0^1z\ln\frac{J_{|m|}^2(j_{|m|n}z)}{J_{|m|}'^{\,2}(j_{|m|n})}dz+\ldots.
\end{equation}
Analysis of the integral in this asymptote shows that, for example, for the rotationally symmetric levels the magnitude of its negative value gets smaller as $n$ grows what makes the entropy $R_{\rho_{n0}}^D(\alpha)$ in this regime an increasing function of the principal quantum index with the ground-state measure being slightly split off from its $n>1$ counterparts that are almost equal to each other.  To find the functional in the opposite limit of the infinite parameter $\alpha$, one employs the relation
\begin{equation}\label{RenyiRelation1}
R_{\rho,\gamma}(\infty)=-\ln\!\left(\!\!\!\begin{array}{c}
\rho_{max}\\
\gamma_{max}
\end{array}\!\!\!\right)
\end{equation}
and the values of the global extrema of the corresponding density found in section~\ref{sec_WaveFunctionsDirichlet}:
\begin{subequations}\label{DirichletRenyiPositionInfinite}
\begin{align}\label{DirichletRenyiPositionInfinite_1}
R_{\rho_{nm}}^D(\infty)&=2\ln a+\ln\pi-2\ln\left|\frac{J_{|m|}(j_{|m|1}')}{J_{|m|}'(j_{|m|n})}\right|.
\intertext{From here, one immediately gets that, at the huge quantum numbers, the R\'{e}nyi entropy unrestrictedly decreases:}
\label{DirichletRenyiPositionInfinite_2}
R_{\rho_{nm}}^D(\infty)&\sim-\ln n,\quad n\rightarrow\infty\\
\label{DirichletRenyiPositionInfinite_3}
R_{\rho_{nm}}^D(\infty)&\sim-\frac{2}{3}\ln|m|,\quad |m|\rightarrow\infty.
\end{align}
\end{subequations}
As at the zero R\'{e}nyi parameter all functionals take the same value of $2\ln a+\ln\pi$, it follows from Equations~\eqref{DirichletRenyiPositionInfinite} that the absolute value of the negative average speed of change of the position entropy increases with $n$ and $|m|$ with its instantaneous value $dR_{\rho_{nm}}^D(\alpha)/d\alpha$ depending additionally on the coefficient $\alpha$ what leads to the intersections in Figure~\ref{Fig_RenyiPositionDirichlet1} of the lines for the different states. Since, as mentioned above, the distinction between the $n>1$ functionals at the tiny $\alpha$ is quite small, the gaps between the individual crossings are not resolved in the scale of this figure. At $\alpha=1$, R\'{e}nyi entropy upon application of the l'H\^{o}pital's rule turns, as expected, into its Shannon counterpart whose nonmonotonic behavior with respect to the indices $n$ and $|m|$, as discussed in section~\ref{sec_ShannonDirichlet}, is explained by the different rate of change of the R\'{e}nyi measure. For the uncertainty relation, it is important to know a position functional at $\alpha=1/2$:
\begin{subequations}\label{DirichletRenyiPositionOneHalf}
\begin{align}\label{DirichletRenyiPositionOneHalf_N}
R_{\rho_{nm}}^D\left(\frac{1}{2}\right)&=2\ln a+\ln\pi+2\ln\left(\frac{2}{\left|J_{|m|+1}(j_{|m|n})\right|}\int_0^1z\left|J_{|m|}(j_{|m|n}z)\right|dz\right).
\intertext{Its particular values for the two lowest $|m|$ read as:}
\label{DirichletRenyiPositionOneHalf_0}
R_{\rho_{n0}}^D\left(\frac{1}{2}\right)&=2\ln a+\ln\pi+2\ln\frac{2[(-1)^{n+1}j_{0n}J_1(j_{0n})-2\sum_{k=1}^{n-1}(-1)^kj_{0k}J_1(j_{0k})]}{j_{0n}^2\left|J_1(j_{0n})\right|}\\
\label{DirichletRenyiPositionOneHalf_1}
R_{\rho_{n1}}^D\left(\frac{1}{2}\right)&=2\ln a+3\ln\pi+2\ln\frac{(-1)^nj_{1n}J_0(j_{1n}){\bf H}_1(j_{1n})+2\sum_{k=1}^{n-1}(-1)^kj_{1k}J_0(j_{1k}){\bf H}_1(j_{1k})}{j_{1n}^2\left|J_2(j_{1n})\right|},
\end{align}
\end{subequations}
where ${\bf H}_\nu(z)$ is Struve function \cite{Abramowitz1}. For later reference, let us specifically single out the ground level:
\begin{equation}\tag{67b$'$}\label{eq:67b'}
R_{\rho_{10}}^D\left(\frac{1}{2}\right)=2\ln a+\ln\pi+2\ln\frac{2}{j_{01}},
\end{equation}
with $\ln\pi+2\ln\frac{2}{j_{01}}=0.77606\ldots$. 

\begin{figure}
\centering
\includegraphics[width=\columnwidth]{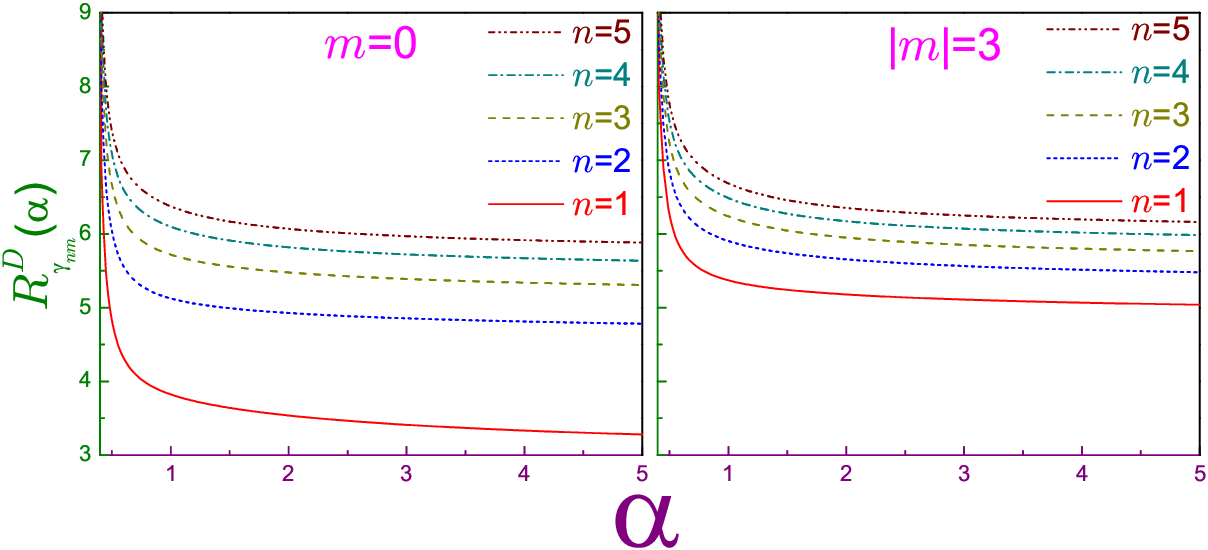}
\caption{\label{Fig_RenyiMomentumDirichlet1}R\'{e}nyi momentum entropy of the Dirichlet disc $R_{\gamma_{nm}}^D(\alpha)$ as a function of the parameter $\alpha$. The same convention as in Figure~\ref{Fig_RenyiPositionDirichlet1} is adopted.}
\end{figure}

A comparison test applied to the improper integral \cite{Fikhtengolts1} in the momentum functionals of the Dirichlet dot
\begin{align}\label{DirichletRenyiMomentum1}
R_{\gamma_{nm}}^D(\alpha)&=-2\ln a+\ln\pi+\frac{1}{1-\alpha}\ln\!\left(\!2j_{|m|n}^{2\alpha}\!\int_0^\infty\!\!\xi\left[\frac{J_{|m|}^2(\xi)}{(j_{|m|n}^2-\xi^2)^2}\right]^\alpha\!\!d\xi\right)\\
\label{DirichletTsallisMomentum1}
T_{\gamma_{nm}}^D(\alpha)&=\frac{1}{\alpha-1}\left(1-\!2\left(\frac{a^2}{\pi}\right)^{\alpha-1}j_{|m|n}^{2\alpha}\!\int_0^\infty\!\!\xi\left[\frac{J_{|m|}^2(\xi)}{(j_{|m|n}^2-\xi^2)^2}\right]^\alpha\!\!d\xi\right)
\end{align}
immediately reveals that it does make sense to talk about the wave vector R\'{e}nyi and Tsallis components at the values of the unitless parameter being only not smaller than
\begin{equation}\label{DirichletThreshold1}
\alpha_{TH}^D=\frac{2}{5},
\end{equation}
which is greater than its 1D counterpart of $1/4$ \cite{Olendski1}. Thus, increasing the dimensionality  from one to two coordinates pushes higher the threshold of the semi-infinite range of the R\'{e}nyi/Tsallis coefficient where the momentum entropies exist. Upon approaching by  $\alpha$ this limiting value, the R\'{e}nyi entropy logarithmically diverges, as is depicted in Figure~\ref{Fig_RenyiMomentumDirichlet1}. Analysis shows that $R_{\gamma_{nm}}^D(\alpha)$ is, at any $\alpha$, a monotonically increasing function of both $|m|$ and $n$, what is inherited by the Shannon functional, as discussed before, section~\ref{sec_ShannonDirichlet}. Its values at the infinite factor are:

for the ground state:
\begin{subequations}\label{DirichletRenyiMomentumInfinite1}
\begin{align}\label{DirichletRenyiMomentumInfinite1_1}
R_{\gamma_{10}}^D(\infty)&=-2\ln a+\ln\pi+2\ln j_{01},
\intertext{as it follows from Equations~\eqref{RenyiRelation1} and~\eqref{Equation1G_Dirichlet}, and}
\label{DirichletRenyiMomentumInfinite1_2}
R_{\gamma_{nm}}^D(\infty)&=-2\ln a+\ln(4\pi)-2\ln|J_{|m|}'(j_{|m|n})|-\frac{6}{j_{|m|n}^2-m^2-5},
\end{align}
\end{subequations}
- for the huge $|m|$ and/or $n$; in particular, in the limit $|m|\rightarrow\infty$ it increases as $\frac{4}{3}\ln|m|$. In Equation~\eqref{DirichletRenyiMomentumInfinite1_1}, $\ln\pi+2\ln j_{01}=2.8996\ldots$, what is the lowest $R_{\gamma_{nm}}^D(\alpha)+2\ln a$ value at any $\alpha$, $n$ and $m$. Similar to the position component, the rate of change $dR_{\gamma_{nm}}^D(\alpha)/d\alpha$ depends on the orbital and coefficient $\alpha$.

\begin{figure}
\centering
\includegraphics[width=0.9\columnwidth]{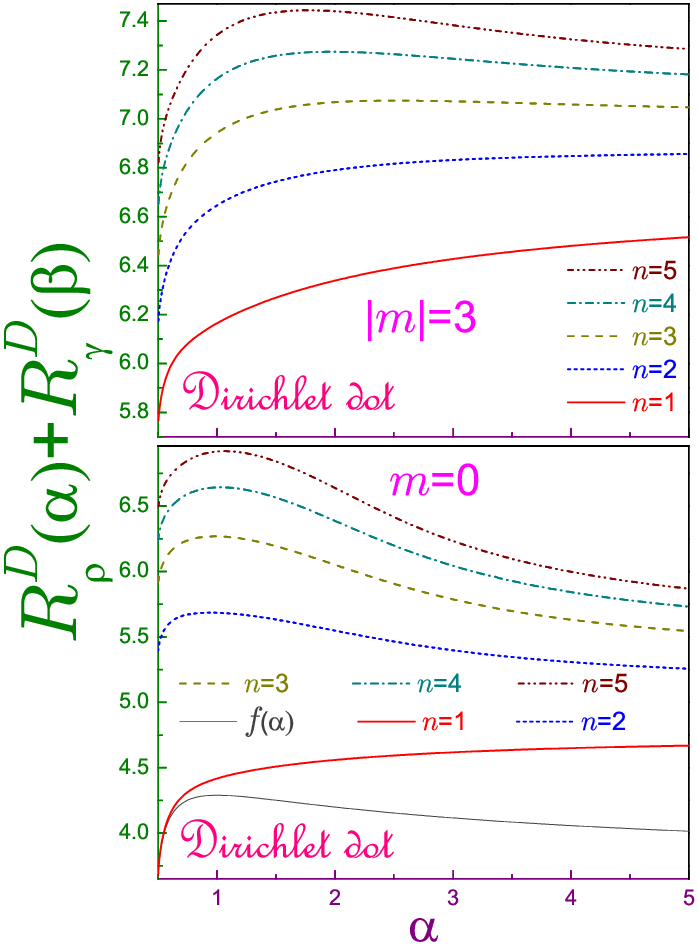}
\caption{\label{Fig_RenyiUncertaintyDirichlet1}Sums of the position and momentum R\'{e}nyi entropies $R_{\rho_{nm}}^D(\alpha)+R_{\gamma_{nm}}^D(\beta)$ of the Dirichlet dot as functions of parameter $\alpha$ with lower panel depicting $m=0$ dependencies and upper window being for $|m|=3$. Solid lines represent $n=1$ orbitals, dotted curves are for $n=2$ states, dashed ones - for $n=3$, dash-dotted dependencies show $n=4$ levels, and dash-dot-dotted ones - $n=5$. In addition, thin solid line in the lower subplot exhibits right-hand side of the R\'{e}nyi uncertainty relation, i.e., function $f(\alpha)$ from Equation~\eqref{Function_f1}. Note different vertical ranges and scales for the two subplots.}
\end{figure}

For the 2D geometry, right-hand side of the R\'{e}nyi uncertainty relation, Equation~\eqref{RenyiUncertainty1}, expressed with the help of the conjugation from Equation~\eqref{RenyiUncertainty2} as a function of the parameter $\alpha$:
\begin{equation}\label{Function_f1}
f(\alpha)=2\left[\ln\pi-\ln\alpha+\frac{\alpha-1/2}{\alpha-1}\ln(2\alpha-1)\right],
\end{equation}
reaches its only maximum of $2(1+\ln\pi)$ at the Shannon regime, $\alpha=1$,
\begin{subequations}\label{HO_Flimits}
\begin{align}
\label{HO_FlimitsOne}
f(\alpha\rightarrow1)&=2(1+\ln\pi)-\frac{1}{3}\,(\alpha-1)^2+\frac{2}{3}\,(\alpha-1)^3+\ldots,
\intertext{and approaches $2\ln2\pi=3.6757\ldots$ at $\alpha=1/2$ and $\alpha=\infty$ \cite{Olendski2,Olendski3}:}
\label{HO_FlimitsOneHalf}
f\left(\alpha\rightarrow\frac{1}{2}\right)&=2\ln2\pi-2[1+\ln(2\alpha-1)](2\alpha-1)+\ldots,\\
\label{HO_FlimitsInfinity}
f(\alpha\rightarrow\infty)&=2\ln2\pi+\frac{\ln2\alpha-1}{\alpha}+\ldots.
\end{align}
\end{subequations}
It is depicted in the lower panel of Figure~\ref{Fig_RenyiUncertaintyDirichlet1} where also several dimensionless sums $R_{\rho_{nm}}^D(\alpha)+R_{\gamma_{nm}}^D(\beta)$ for $m=0$ and $|m|=3$ are plotted too. It shows that, similar to the previous geometries \cite{Olendski1,Olendski2,Olendski3}, at $\alpha=1/2$ (i.e., at $\beta=\infty$, as it follows from Equation~\eqref{RenyiUncertainty2}) the lowest-energy orbital converts the R\'{e}nyi uncertainty into the identity, as also a comparison between Equations~\eqref{eq:67b'}, \eqref{DirichletRenyiMomentumInfinite1_1}  and \eqref{HO_FlimitsOneHalf} confirms; namely, at this parameter $\beta$, the momentum item in the left-hand side of Equation~\eqref{RenyiUncertainty1} becomes:
\begin{subequations}\label{RenyiProof1}
\begin{align}\label{RenyiProof1_Momentum1}
R_{\gamma_{nm}}(\beta)|_{\alpha=1/2}&=-2\ln|\Phi_{nm}|_{max},
\intertext{where for the angle-independent, $m=0$, level of the ground band, $n=1$, the global extremum is achieved at the zero momentum, as it has been found in section~\ref{sec_WaveFunctionsDirichlet}:}
\label{RenyiProof1_Momentum2}
R_{\gamma_{10}}(\beta)|_{\alpha=1/2}&=-2\ln\Phi_{10}(0).
\intertext{On the other hand, the position item in the same regime reads:}
\label{RenyiProof1_Position1}
R_{\rho_{nm}}(\alpha)|_{\alpha=1/2}&=2\ln\int_{{\cal D}_\rho^2}|\Psi_{nm}({\bf r})|d{\bf r},
\intertext{where the general definition, Equation~\eqref{Renyi1_R}, has been used. In this last equation, the absolute value of the position function is equal at any $\bf r$ to the function itself only for the rotationally symmetric (what means that $\Psi_{n0}$ is real) dependence of the lowest band, $n=1$, when the radial function does not change sign along the whole interval $0\leq r<a$, cf. Figure~\ref{Fig_FunctionsDirichletR1}. Then, for this level, as it follows from the definition of the Fourier transform, Equation~\eqref{Fourier1_1}:}
\label{RenyiProof1_Position2}
R_{\rho_{10}}(\alpha)|_{\alpha=1/2}&=2\ln\left(2\pi\Phi_{10}(0)\right).
\end{align}
\end{subequations}
Adding Equations~\eqref{RenyiProof1_Position2} and \eqref{RenyiProof1_Momentum2} yields the lower value of the function $f(\alpha)$, Equation~\eqref{HO_FlimitsOneHalf}. For any other orbital, R\'{e}nyi uncertainty relation is a strict inequality at any parameter $\alpha$, as Figure~\ref{Fig_RenyiUncertaintyDirichlet1} demonstrates. In the opposite limit of the infinitely large coefficient $\alpha$ the value of $R_{\gamma_{nm}}(1/2)$ can be calculated numerically only what, combined with the analytic expression of the position component, Equation~\eqref{DirichletRenyiPositionInfinite_1}, yields:
$$\left.\left[R_{\rho_{10}}^D(\alpha)+R_{\gamma_{10}}^D(\beta)\right]\right|_{\alpha=\infty}=4.665\ldots$$
$$\left.\left[R_{\rho_{20}}^D(\alpha)+R_{\gamma_{20}}^D(\beta)\right]\right|_{\alpha=\infty}=5.026\ldots$$
etc. Calculations show that the sum in the uncertainty relation is at any R\'{e}nyi coefficient an increasing function of both $n$ and $|m|$. An interplay between the opposite contributions of the first and second items regulates the shape of the corresponding curve: if, with the factor $\alpha$ raising from one half, the decline in $R_{\rho_{nm}}(\alpha)$ can not compensate simultaneous growth of $R_{\gamma_{nm}}\left(\frac{\alpha}{2\alpha-1}\right)$, the left-hand side of the R\'{e}nyi inequality is a monotonically increasing function in the whole range $[+1/2,\infty)$; however, if, at some extreme $\alpha$, the absolute value of the deduction from the position component begins to overweigh the positive donation from the momentum functional, one observes the sole maximum in the sum dependence on $\alpha$ whose magnitude, width and location are determined by the corresponding orbital. As shown above, the rate of change $dR_{\rho_{nm}}^D(\alpha)/d\alpha$ at the fixed $m$ is the smallest for the ground band; accordingly, the sum $R_{\rho_{1m}}(\alpha)+R_{\beta_{1m}}(\beta)$ at any magnetic index does not exhibit the extemum. As the lower panel of Figure~\ref{Fig_RenyiUncertaintyDirichlet1} demonstrates, all other rotationally symmetric levels do possess the maximum that, due to Equation~\eqref{DirichletRenyiPositionInfinite_2}, grows more conspicuous with $n$. For the first excited band, $n=2$, the maximum disappears at $|m|\geq3$; and for $n=3$ - at $|m|\geq5$, etc.

\begin{figure}
\centering
\includegraphics[width=0.9\columnwidth]{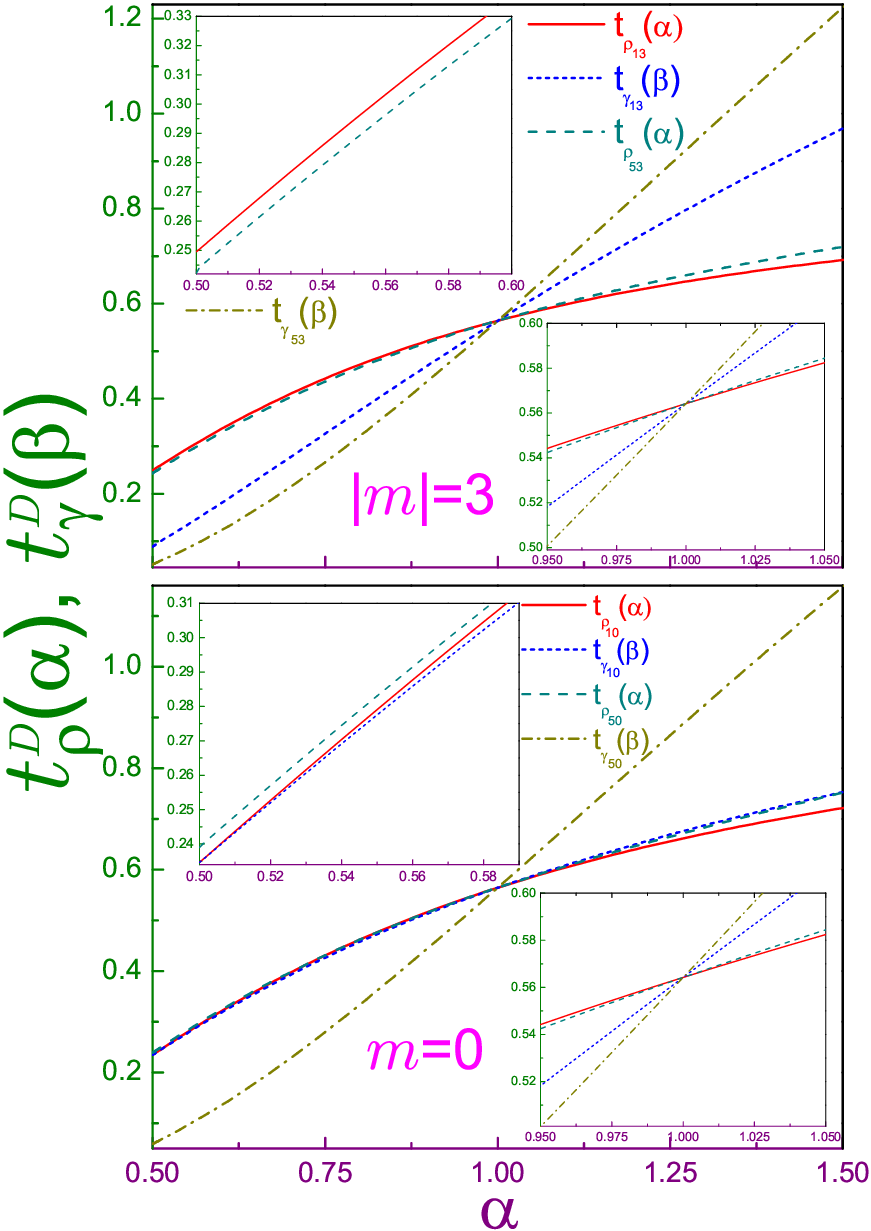}
\caption{\label{Fig_TsallisUncertaintyDirichlet1}Dimensionless Tsallis position $t_{\rho_{nm}}^D(\alpha)$ and momentum $t_{\gamma_{nm}}^D(\beta)$ functions, Equations~\eqref{TsallisInequalityDirichlet1}, in terms of parameter $\alpha$ for the Dirichlet rotationally symmetric (lower window) and $|m|=3$ (upper panel) orbitals where solid and dashed lines stand for the position parts of $n=1$ and $n=5$ states, respectively, whereas dotted and dash-dotted curves depict momentum dependencies of the same levels. In each subplot, the lower right insets exhibit functions in $\alpha=1$ vicinity and upper left ones show $t_{\rho_{1m}}^D(\alpha)$ and $t_{\gamma_{1m}}^D(\beta)$ parameters close to the left border $\alpha=1/2$.}
\end{figure}

Expression for the Tsallis entropy, written for any \textit{continuous} probability distribution, as those, e.g., in Equations~\eqref{Tsallis1}, suffers a dimensionality problem what is further exemplified in Equation~\eqref{DirichletTsallisMomentum1}: namely, it is a sum of two items the first one being a \textit{dimensionless} unity whereas the second term is the length raised to the power $d(\alpha-1)$ or its inverse. Accordingly, in this case one can not use the Tsallis entropy directly. However, the corresponding uncertainty relation, Equation~\eqref{TsallisInequality1}, or, equivalently, Equation~\eqref{Sobolev1}, can be analyzed; namely, they take the form:
\begin{equation}\label{TsallisInequality2}
a^\frac{1-\alpha}{\alpha}t_{\rho_{nm}}(\alpha)\geq a^\frac{1-\beta}{\beta}t_{\gamma_{nm}}(\beta),
\end{equation}
where \textit{dimensionless} position $t_\rho(\alpha)$ and wave vector $t_\gamma(\beta)$ functions for the Dirichlet BC read:
\begin{subequations}\label{TsallisInequalityDirichlet1}
\begin{align}\label{TsallisInequalityDirichlet1_Position}
t_{\rho_{nm}}^D(\alpha)&=\frac{1}{\pi^{1/2}}\left(2\alpha\int_0^1z\!\left[\frac{J_{|m|}^2(j_{|m|n}z)}{J_{|m|+1}^2(j_{|m|n})}\right]^\alpha\!\!dz\right)^\frac{1}{2\alpha}\\
\label{TsallisInequalityDirichlet1_Momentum}
t_{\gamma_{nm}}^D(\beta)&=\frac{1}{\pi^{1/2}}\left(2\beta\int_0^\infty\!\!\xi\left[\frac{J_{|m|}^2(\xi)}{(j_{|m|n}^2-\xi^2)^2}\right]^\beta\!\!d\xi\right)^\frac{1}{2\beta}.
\end{align}
\end{subequations}
They are shown in Figure~\ref{Fig_TsallisUncertaintyDirichlet1} for $m=0$ (lower window) and $|m|=3$ (upper subplot) and $n=1,5$ states. Obviously, all the dependencies cross at $\alpha=1$ where their values regardless of the BC type are
\begin{equation}\label{TsallisInequality3}
t_{\rho_{nm}}(1)=t_{\gamma_{nm}}(1)=\frac{1}{\pi^{1/2}}=0.5641\ldots,
\end{equation}
and Tsallis uncertainty relation, Equations~\eqref{Sobolev1}, \eqref{TsallisInequality1} and \eqref{TsallisInequality2}, holds true to the left of it only. In addition, using the same arguments as for the R\'{e}nyi entropies, one finds that the lowest, $n=1$, $m=0$, level equalizes position and momentum dependencies also at $\alpha=1/2$ where they are
\begin{equation}\label{TsallisInequality4}
t_{\rho_{10}}\left(\frac{1}{2}\right)=t_{\gamma_{10}}\left(\infty\right)=\frac{1}{a}\Phi_{10}({\bf 0}),
\end{equation}
what for the Dirichlet dot, according to Equation~\eqref{DirichletFunction1_Momentum}, is
$$\frac{1}{a}\Phi_{10}^D({\bf 0})=\frac{1}{\pi^{1/2}j_{01}}=0.2346\ldots.$$

\begin{figure}
\centering
\includegraphics[width=\columnwidth]{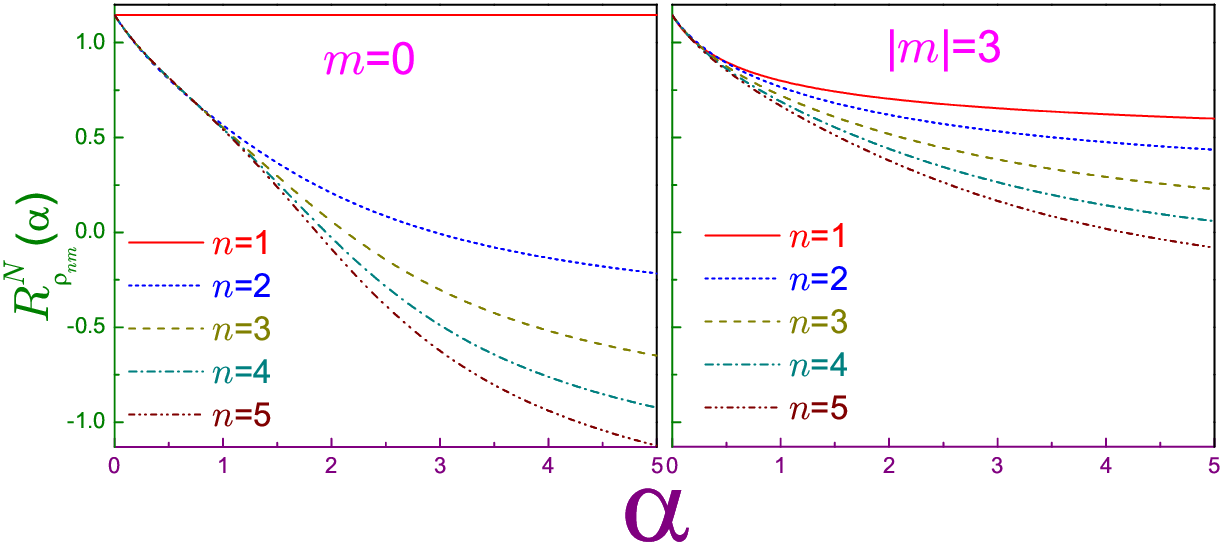}
\caption{\label{Fig_RenyiPositionNeumann1}The same as in Figure~\ref{Fig_RenyiPositionDirichlet1} but for the Neumann dot.}
\end{figure}

\subsection{Neumann dot}\label{sec_RenyiNeumann}
Similar to the quantum well \cite{Olendski1}, R\'{e}nyi position component of the Neumann ground state does not depend on the coefficient $\alpha$, whereas, contrary to the 1D geometry, all other levels have their own distinct $R_{\rho_{nm}}^N(\alpha)$ characteristics:
\begin{equation}\label{NeumannRenyiPosition1}
R_{\rho_{nm}}^N(\alpha)=2\ln a+\ln\pi+\left\{\begin{array}{cl}
0,&n=1,\,m=0\\
\frac{1}{1-\alpha}\ln\!\left(\!2\!\left(\frac{j_{|m|n}'^{\,2}}{j_{|m|n}'^{\,2}-m^2}\right)^\alpha\!\!\mathop{\mathlarger{\int}}_{\!\!\!0}^1z\!\left[\frac{J_{|m|}^2(j_{|m|n}'z)}{J_{|m|}^2(j_{|m|n}')}\right]^\alpha\!\!dz\right),&{\rm all\,\,other\,\,cases}
\end{array}
\right.,
\end{equation}
as Figure~\ref{Fig_RenyiPositionNeumann1} demonstrates. For the extremely large R\'{e}nyi parameter, one gets:
\begin{equation}\label{NeumannRenyiPositionInfinite}
R_{\rho_{nm}}^N(\infty)=2\ln a+\ln\pi-2\ln\left|\frac{j_{|m|n}'}{(j_{|m|n}'^{\,2}-m^2)^{1/2}}\frac{J_{|m|}(j_{|m|1}')}{J_{|m|}(j_{|m|n}')}\right|,
\end{equation}
which means that in the limit of one and/or two huge quantum numbers this value basically coincides with its Dirichlet counterparts, Equations~\eqref{DirichletRenyiPositionInfinite_2} and \eqref{DirichletRenyiPositionInfinite_3}. Explanation of this is similar to the measures discussed above, section~\ref{sec_ShannonNeumann}, and 1D case \cite{Olendski1}: at the huge energies, the corresponding orbital does not tell one BC type from the other. Also, similar to Figure~\ref{Fig_RenyiPositionDirichlet1}, the scale of Figure~\ref{Fig_RenyiPositionNeumann1} does not resolve different line crossings.

\begin{figure}
\centering
\includegraphics[width=\columnwidth]{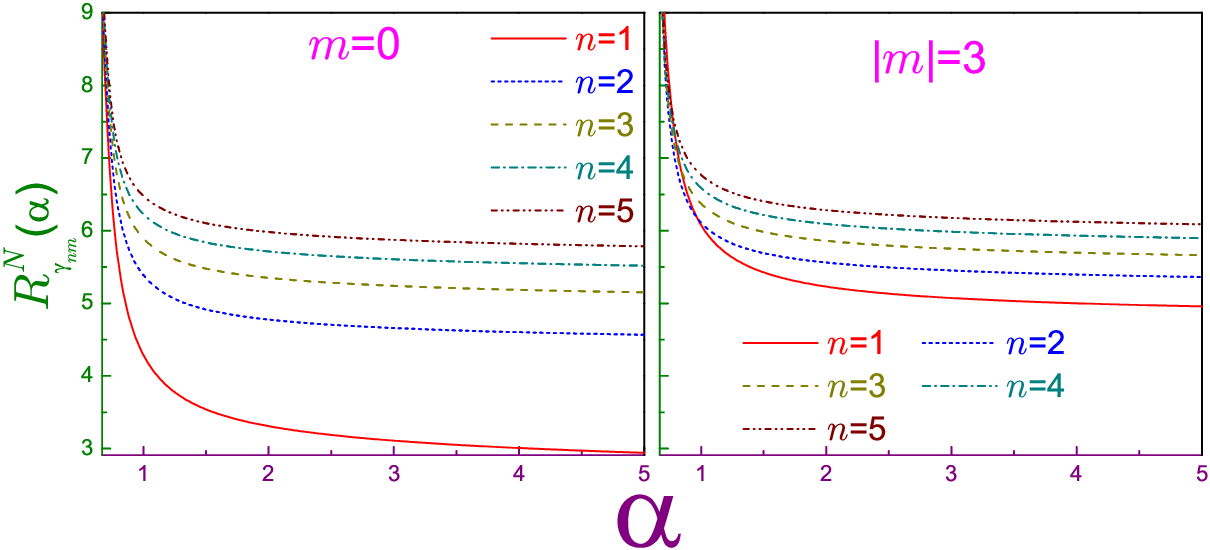}
\caption{\label{Fig_RenyiMomentumNeumann1}The same as in Figure~\ref{Fig_RenyiMomentumDirichlet1} but for the Neumann dot.}
\end{figure}

Consideration of the momentum components, for example, the R\'{e}nyi one,
\begin{equation}\label{NeumannRenyiMomentum1}
R_{\gamma_{nm}}^N(\alpha)=-2\ln a+\ln\pi+\frac{1}{1-\alpha}\ln\!\left(2\!\left(\frac{j_{|m|n}'^{\,2}}{j_{|m|n}'^{\,2}-m^2}\!\!\right)^\alpha\!\!\!\!\int_0^\infty\!\!\!\xi\left[\frac{\xi^2}{(j_{|m|n}'^{\,2}-\xi^2)^2}J_{|m|}'^{\,2}(\xi)\right]^\alpha\!\!d\xi\right),
\end{equation}
and, in particular, the ground-state functional:
\begin{equation}\tag{81$'$}\label{eq:81'}
R_{\gamma_{10}}^N(\alpha)=-2\ln a+\ln\pi+\frac{1}{1-\alpha}\ln\int_0^\infty\xi^{1-2\alpha}\left[J_1^2(\xi)\right]^\alpha d\xi,
\end{equation}
makes it clear that they are defined at the corresponding coefficient only lying above
\begin{equation}\label{NeumannThreshold1}
\alpha_{TH}^N=\frac{2}{3}.
\end{equation}
Note that this threshold is greater than both its 2D Dirichlet counterpart, Equation~\eqref{DirichletThreshold1}, and the 1D Neumann one of $1/2$ \cite{Olendski1}. Thus, a transition from the 1D to the 2D geometry increases both brinks in such a way that the Neumann verge remains higher than the Dirichlet one. Another difference between the two BCs is the fact that at any non-zero azimuthal number the momentum entropy $R_{\gamma_{nm}}^N(\alpha)$ looses its monotonic dependence on the band index as the R\'{e}nyi coefficient tends to the critical value from Equation~\eqref{NeumannThreshold1}. Right panel of Figure~\ref{Fig_RenyiMomentumNeumann1} shows the corresponding intersections of the curves for different $n$ whereas the functionals for the rotationally symmetric orbitals continue to obey the rule $R_{\gamma_{n0}}^N(\alpha)<R_{\gamma_{n+1,0}}^N(\alpha)$ in the whole range $[\alpha_{TH}^N,+\infty)$ as was the case for all Dirichlet levels with arbitrary $n$ and $m$. Crossing of the entropies of the two neighboring bands shifts to the left as the principal quantum number increases and the greater azimuthal index $|m|$ pushes them to the higher $\alpha$. This nonmonotonicity was already noticed before for the Shannon entropies, section~\ref{sec_ShannonNeumann}. At infinity, the functionals become:

for the ground state:
\begin{subequations}\label{NeumannRenyiMomentumInfinite1}
\begin{align}\label{NeumannRenyiMomentumInfinite1_1}
R_{\gamma_{10}}^N(\infty)&=-2\ln a+\ln4\pi,
\intertext{with $\ln4\pi=2.5310\ldots$, which is smaller than for the Dirichlet BC; and for the huge $|m|$ and/or $n$:}
R_{\gamma_{nm}}^N(\infty)&=-2\ln a+\ln(4\pi)\!-\!\ln\!\!\left(\!\left[1\!-\!\left(\frac{m}{j_{|m|n}'}\right)^2\right]\!\!J_{|m|}^2\!\left(j_{|m|n}'\right)\!\!\right)\nonumber\\
\label{NeumannRenyiMomentumInfinite1_2}
&-\frac{6m^4}{\left(j_{|m|n}'^{\,2}-m^2\right)\left[j_{|m|n}'^{\,4}-2m^2j_{|m|n}'^{\,2}+m^2(m^2+5)\right]},
\end{align}
\end{subequations}
what essentially is indistinguishable from the other edge condition.

\begin{figure}
\centering
\includegraphics[width=0.9\columnwidth]{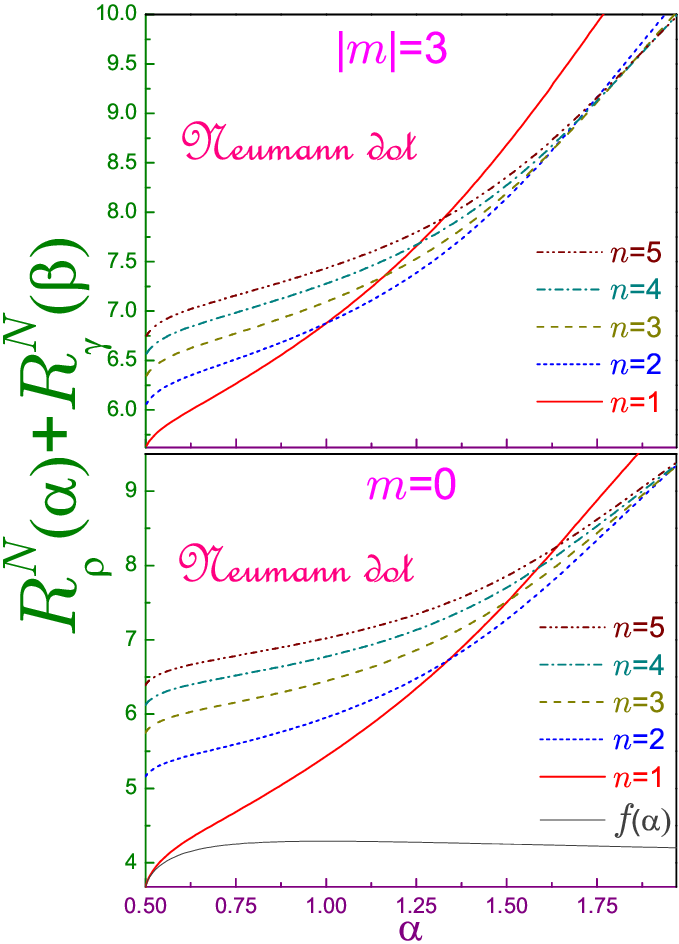}
\caption{\label{Fig_RenyiUncertaintyNeumann1}The same as in Figure~\ref{Fig_RenyiUncertaintyDirichlet1} but for the Neumann dot.}
\end{figure}

Neumann threshold from Equation~\eqref{NeumannThreshold1} has a drastic effect on the uncertainty relation; namely, since it is greater than one half, it follows from the conjugation relation, Equation~\eqref{RenyiUncertainty2}, that the maximal value of the R\'{e}nyi coefficient $\alpha$ in $R_{\rho_{nm}}^N(\alpha)+R_{\gamma_{nm}}^N\left(\frac{\alpha}{2\alpha-1}\right)$ is just two, contrary to all other considered so far Dirichlet and Neumann geometries \cite{Olendski1} where it varied in the whole range $[1/2,+\infty)$. This is depicted in Figure~\ref{Fig_RenyiUncertaintyNeumann1} that shows a monotonic growth of the R\'{e}nyi sum of each orbital and its logarithmic divergence as the R\'{e}nyi factor $\alpha$ approaches the value of two and, accordingly, the conjugated coefficient $\beta$ tends to the limit from Equation~\eqref{NeumannThreshold1}. Similar to all previously discussed structures \cite{Olendski1,Olendski2,Olendski3}, the ground state tightens the uncertainty relation at $\alpha=1/2$. Explanation of this is the same as for the Dirichlet disc, section~\ref{sec_RenyiDirichlet}. Another remarkable feature specific to the non-Dirichlet BC are intersections of the sums for the different bands: the total entropy for the fixed $m$ that at $\alpha=1/2$ is arranged in the increasing order of the principal index looses its $n$ monotonicity with the growth of the R\'{e}nyi coefficient; for example, the ground state crosses its $n=1$ counterpart at $\alpha\sim1.34$. For this particular case, one has to recall that the Neumann ground-state position entropy is independent of $\alpha$ and the decrease of all other $R_{\rho_{n0}}^N(\alpha)$ at $\alpha\gtrsim1$ steepens for the higher $n$, as Figure~\ref{Fig_RenyiPositionNeumann1} shows.  This makes the speed of growth of $R_{\rho_{n0}}^N(\alpha)+R_{\gamma_{n0}}^N(\beta)$ a monotonically decreasing function of the band resulting in the corresponding crossings that for the increasing number $n$ lie closer to $\alpha=2$. For the $m\neq0$ states, one has to recall that their momentum functionals exhibit intersections per se, see Figure~\ref{Fig_RenyiMomentumNeumann1}, and the addition of the position part just shifts them to the left, as compared to rotationally symmetric case. Physically, it means that the Neumann BC makes the total amount of information $R_\rho(\alpha)+R_\gamma(\beta)$ a much  more sensitive function of the R\'{e}nyi coefficient; in particular, in contrast to the Dirichlet requirement, one can collect, depending on $\alpha$, less knowledge about the lower lying states.

Since the overall picture of the Tsallis uncertainty relation of the Neumann BC in the range from Equation~\eqref{Sobolev2} is qualitatively identical to that of the Dirichlet one, we do not exhibit the corresponding plots here. Quantitatively, the magnitude of the position and wave vector parts of the lowest-energy state at the left edge of this interval where they are equal to each other is$$\frac{1}{a}\Phi_{10}^N({\bf 0})=\frac{1}{2\pi^{1/2}}=0.2820\ldots,$$which lies above its Dirichlet counterpart, section~\ref{sec_RenyiDirichlet}.

\section{Conclusions and future outlook}\label{sec_Conclusions}
As a first step in a comparative exploration of the information measures of the 2D Dirichlet and Neumann quantum dots, the formulas for the orthonormalized position waveforms, Equations~\eqref{DirichletFunction1_Position} and \eqref{NeumannPositionFunction1}, have been supplemented by the analytic expressions of their momentum counterparts, Equations~\eqref{DirichletFunction1_Momentum} and \eqref{NeumannMomentumFunction1}. Armed with this knowledge, one can efficiently compute in either space the Shannon, R\'{e}nyi and Tsallis entropies together with the Fisher informations and Onicescu energies. As a result, it was shown that the distinction between the corresponding quantum information measures for the different surface requirements decreases at the higher indexes. 2D geometry lifts 1D generacy of the position Shannon, Onicescu and R\'{e}nyi functionals what for the latter one is especially vividly seen at the moderate and stong parameters $\alpha$. It was discovered that the lowest threshold of the semi-infinite range of the definition of the momentum R\'{e}nyi/Tsallis entropies is not only BC- but also dimensionality-dependent too; namely, a switching from the 1D well to the 2D disc increases both critical values from $1/4$ and $1/2$ to $2/5$ and $2/3$, respectively, i.e., in such a way that the Dirichlet one continues to stay below its Neumann fellow. This distinction has its drastic consequence in different behavior of the R\'{e}nyi sum $R_\rho(\alpha)+R_\gamma\left(\frac{\alpha}{2\alpha-1}\right)$ that enters corresponding entropic inequality, Equation~\eqref{RenyiUncertainty1}; viz., the Dirichlet coefficient $\alpha$ runs through the semi-infinite range $[1/2,+\infty)$  and Neumann counterpart can vary between one half and two only with the logarithmic divergence of the sum at the right edge. Similar to all other structures \cite{Olendski1,Olendski2,Olendski3}, it has been confirmed that the lowest-energy orbital at the left rim of these intervals tightens both R\'{e}nyi and Tsallis inequalities.

Based on the measures discussed above, miscellaneous complexities are defined: $e^SO$ \cite{Catalan1}, $\frac{1}{2\pi e}e^{2S/d}I$ \cite{Dembo1,Vignat1}, $e^{R(\alpha)}O$ \cite{Antolin1,Nath1} and others, see References \cite{Sen1,Toranzo1} and literature therein. Each of them finds its applications in different fields. To save space, they have not been discussed here but this can be easily done using our results; for example, it is known that for any $d$-dimensional space neither position nor momentum component of the first product can never be less than unity, $e^SO\geq1$ \cite{LopezRosa1}, with the equality being reached only for the uniform distribution with a finite volume support; indeed, above relation is tightened by the position component of the lowest Neumann orbital, as it immediately follows from Equations~\eqref{NeumannAnalytic1_1} and \eqref{NeumannAnalytic1_3}. Picking up corresponding Dirichlet, Tables~\ref{Table_EntropyDirichlet} and \ref{Table_OnicescuDirichlet}, and Neumann, Tables~\ref{Table_EntropyNeumann} and \ref{Table_OnicescuNeumann}, entries, one sees that in all other cases the equation above in this paragraph takes the form of a strict inequality.

There are three obvious extensions of the present research. First, at the moment the table of the critical magnitudes determining the lower thresholds of the momentum R\'{e}nyi/Tsallis entropies looks as follows:
$$\begin{array}{ccc}
&{\rm Dirichlet}&{\rm Neumann}\\
{\rm 1D}&1/4&1/2\\
{\rm 2D}&2/5&2/3
\end{array}.
$$
It is natural to wonder: if a transition from the 1D to the 2D geometry increases both thresholds of the momentum functionals, then what are they for the spherical quantum dot and, even more generally, for the $d$-dimensional domain with $d\geq4$? For the 3D geometry with ${\bf r}_{3D}\equiv(r,\theta_{\bf r},\varphi_{\bf r})$ and ${\bf k}_{3D}\equiv(k,\theta_{\bf k},\varphi_{\bf k})$, the waveforms are:

for the Dirichlet BC:
\begin{subequations}\label{Function3D}
\begin{align}\label{Function3D_PositionDirichlet}
3D:&\Psi_{nlm}^D(r,\theta_{\bf r},\varphi_{\bf r})=\frac{2^{1/2}}{a^{3/2}j_{l+1}^{(s)}(j_{l+1/2,n})}j_l^{(s)}\!\left(j_{l+1/2,n}\frac{r}{a}\right)Y_{lm}(\theta_{\bf r},\varphi_{\bf r})\\
\label{Function3D_MomentumDirichlet}
3D:&\Phi_{nlm}^D(k,\theta_{\bf k},\varphi_{\bf k})=(-i)^la^{3/2}\frac{2}{\pi^{1/2}}\frac{j_{l+1/2,n}}{j_{l+1/2,n}^2-(ak)^2} j_l^{(s)}(ak)Y_{lm}(\theta_{\bf k},\varphi_{\bf k});\\
\intertext{for the Neumann requirement:}
\label{Function3D_PositionNeumann}
3D:&\Psi_{nlm}^N(r,\theta_{\bf r},\varphi_{\bf r})=\frac{2^{1/2}}{a^{3/2}}\frac{a_{l,n}'}{\left[a_{l,n}'^{\,2}-l(l+1)\right]^{1/2}}\frac{j_l^{(s)}\!\left(a_{l,n}'\frac{r}{a}\right)}{j_l^{(s)}(a_{l,n}')}Y_{lm}(\theta_{\bf r},\varphi_{\bf r})\\
\label{Function3D_MomentumNeumann}
3D:&\Phi_{nlm}^N(k,\theta_{\bf k},\varphi_{\bf k})=(-i)^la^{3/2}\frac{2}{\pi^{1/2}}\frac{a_{l,n}'}{\left[a_{l,n}'^{\,2}-l(l+1)\right]^{1/2}}\frac{ak}{a_{l,n}'^{\,2}-(ak)^2}{j_l^{(s)}}'(ak)Y_{lm}(\theta_{\bf k},\varphi_{\bf k}).
\end{align}
\end{subequations}
Here, $l=0,1,2,\ldots$ is the azimuthal quantum number and the integer magnetic index $m$ changes in the range from $-l$ to $+l$, $Y_{lm}(\theta,\varphi)$ are standard  orthonormalized spherical harmonics \cite{Morse2,Dicke1,Messiah1}, $j_l^{(s)}(z)=\sqrt{\pi/(2z)}J_{l+1/2}(z)$ is the spherical Bessel function \cite{Abramowitz1}, and $a_{l,n}'$ is the value at which its derivative turns to zero, ${j_l^{(s)}}'(a_{l,n}')=0$ \cite{Abramowitz1}. Momentum dependencies from Equations~\eqref{Function3D_MomentumDirichlet} and \eqref{Function3D_MomentumNeumann} were obtained from the general definition, Equation~\eqref{Fourier1_1}, and the well-known plane-wave expansion \cite{Messiah1}:
$$e^{i{\bf kr}}=4\pi\sum_{l=0}^\infty\sum_{m=-l}^li^lj_l^{(s)}(kr)Y_{lm}(\theta_{\bf r},\varphi_{\bf r})Y_{lm}^*(\theta_{\bf k},\varphi_{\bf k}).$$
Then, without detailed calculations of the entropies but simply applying a convergence test \cite{Fikhtengolts1} to the corresponding functionals from Equations~\eqref{Renyi1_K} and \eqref{Tsallis1_K} with the momentum waveforms from Equations~\eqref{Function3D_MomentumDirichlet} and \eqref{Function3D_MomentumNeumann}, one supplements the table above in this paragraph by the line:
$$\begin{array}{ccc}
{\rm 3D}&1/2&3/4
\end{array},
$$
what immediately leads to the conjecture that for the $d$-dimensional structure the Dirichlet and Neumann momentum thresholds are, respectively: 
\begin{subequations}\label{Ddimensional1}
\begin{align}\label{Ddimensional1_D}
\alpha_{TH}^D(d)&=\frac{d}{d+3}\\
\label{Ddimensional1_N}\alpha_{TH}^N(d)&=\frac{d}{d+1}.
\end{align}
\end{subequations}
A proof of the correctness of Equations~\eqref{Ddimensional1}, which employs hyperspherical Bessel functions and harmonics \cite{Avery1}, and their influence on the R\'{e}nyi and Tsallis uncertainty relations will be presented elsewhere \cite{Olendski7}. Second, if one considers a 2D annulus with the inner $r_0$ and outer $r_0+a$ radii, then at the vanishing $r_0$ the results presented above should be recovered and at $r_0\rightarrow\infty$ some analog of the 1D structure is approached and since in both these limits the momentum thresholds are different, it would be interesting to see their transformation with the ring varying from the thick (small $r_0$) to the thin (huge $r_0$) one. Another exciting generalization is the influence of the transverse magnetic field $\bf B$ when its increasing magnitude pushes the position densities to the dot center thus diminishing the confinement influence and converting the orbitals into the Landau levels what means that its strong values should suppress both the Dirichlet and Neumann thresholds to zero wiping out in this way the difference between them. Position waveforms and energy spectra of both Dirichlet \cite{Geerinckx1,Lent1,Constantinou1,Olendski6} and Neumann \cite{Olendski6,Constantinou2} disc in the field $B$ are known but apparently nobody considered momentum parts and associated quantum-information measures in both spaces.
  
\section{Acknowledgements}
Research was supported by Competitive Research Project No. 2002143087 from the Research Funding Department, Vice Chancellor for Research and Graduate Studies, University of Sharjah.

\bibliographystyle{}

\end{document}